\documentclass{JHEP3}
\usepackage{epsfig}
\def\eslt{\not\!\!{E_T}}

\def\to{\rightarrow}

\def\bi{\begin{itemize}}
\def\ei{\end{itemize}}
\def\be{\begin{equation}}
\def\ee{\end{equation}}
\def\bea{\begin{eqnarray}}
\def\eea{\end{eqnarray}}

\def\tg{\tilde g}

\def\tw{\widetilde W}
\def\tz{\widetilde Z}
\def\alt{\stackrel{<}{\sim}}

\title{
Model Independent Approach to Focus Point Supersymmetry:
from Dark Matter to Collider Searches}

\author{Howard Baer$^a$, Tadas Krupovnickas$^b$, Stefano Profumo$^a$ and Piero Ullio$^c$\\ 

$^a$ Department of Physics, Florida State University, Tallahassee, FL 32306, USA\\
$^b$ High Energy Theory Group, Brookhaven National Laboratory, Upton, NY 11973, USA\\
$^c$ SISSA/ISAS, via Beirut 2-4, 34013 Trieste, Italy\\

E-mail: \email{baer@hep.fsu.edu}, \email{tadas@quark.phy.bnl.gov}, \email{profumo@hep.fsu.edu}, \email{ullio@he.sissa.it}}
\preprint{\vbox{\hbox{FSU-HEP-050730, BNL-HET-05/21} }}

\abstract{
The focus point region of supersymmetric models is compelling in that it simultaneously features low fine-tuning, provides a decoupling solution to
the SUSY flavor and CP problems, suppresses proton decay rates and can accommodate the WMAP measured cold dark matter (DM) relic density
through a mixed bino-higgsino dark matter particle. We present the focus point region in terms of a weak scale parameterization, which allows
for a relatively model independent compilation of phenomenological constraints and prospects. We present direct and indirect 
neutralino dark matter detection rates for two different halo density profiles, and show that prospects for direct DM detection and
indirect detection via neutrino telescopes such as IceCube and anti-deuteron searches by GAPS are especially promising. 
We also present LHC reach prospects via
gluino and squark cascade decay searches, and also via clean trilepton signatures arising from chargino-neutralino
production. Both methods provide a reach out to $m_{\tg}\sim 1.7$ TeV. At a TeV-scale linear $e^+e^-$ collider (LC), the maximal
reach is attained in the $\tz_1\tz_2$ or $\tz_1\tz_3$ channels. In the DM allowed region of parameter space, a $\sqrt{s}=0.5$ TeV
LC has a reach which is comparable to that of the LHC. 
However, the reach of a 1 TeV LC extends out to $m_{\tg}\sim 3.5$ TeV.
}


\begin{document}

\section{Introduction}
\label{sec:intro}
Supersymmetric models of particle physics provide a compelling case for physics beyond the Standard Model (SM).
However, in spite of their many successes, they also provide a long list of potential problems. For instance,
supersymmetric models are supposed to provide a solution to the fine-tuning problem which arises when the SM is embedded
in theories with high mass scales beyond a few TeV. However, the generally excellent agreement of precision EW
observables with SM predictions, along with null search results for various rare decays and other loop induced processes, 
points to a rather heavy sparticle mass spectrum, with sparticles 
typically
in the TeV regime. These observations are supported by recent search results from LEP2 that the chargino mass
$m_{\tw_1}>103.5$ GeV and the SM Higgs mass $m_{H_{SM}}>114.4$ GeV. The latter limit, when applied to the
Higgs bosons of the MSSM, also implies relatively heavy top squarks. A rather
heavy sparticle mass spectrum, on the other hand, naively seems to re-introduce the fine-tuning problem into supersymmetric models.

In addition, in the 124 parameter Minimal Supersymmetric Standard Model (MSSM), unsuppressed FCNC effects arise
from general Lagrangian parameters, as do large contributions to the electric dipole moments of the electron and neutron
from possibly large $CP$-violating phases. It has been noted by many authors that scalar  masses in the multi-TeV
regime can act to suppress most or all of these undesired effects to within levels tolerated by data\cite{decouple}. 
Multi-TeV scalar masses can be reconciled with a no-fine-tuning requirement in two cases. In one case, 
inverted scalar mass hierarchy models\cite{imh} (IMH) require scalars of the first two generations to be in the multi-TeV regime, while
scalars of the third generation, which enter fine-tuning calculations, remain at sub-TeV levels. In practice, the radiatively driven IMH
models require $t-b-\tau$ Yukawa coupling unification and non-universal soft SUSY breaking (SSB) Higgs masses
to be viable\cite{feng,yuk}. The second case, the topic of this paper, 
is that of hyperbolic branch\cite{ccn} or focus point models\cite{fmm} (HB/FP), wherein all three
generations of scalars can be in the multi-TeV regime, while fine-tuning is respected for low values
of the GUT scale universal gaugino mass $m_{1/2}$.

The HB/FP region appears already in Ref. \cite{bcpt} as a region in the 
$m_0\ vs.\ m_{1/2}$ plane of the minimal supergravity (mSUGRA)
model where $m_0$ is in the multi-TeV regime, but where the absolute value of the superpotential $\mu$ parameter becomes small,
adjacent to regions where radiative electroweak symmetry breaking (REWSB) fails to occur (where $\mu^2<0$).
The small value of $|\mu |$ leads to a mixed higgsino-bino LSP (a mixed higgsino dark matter (DM) candidate) 
and a rather light spectrum of charginos and neutralinos. This mixed higgsino region with large scalar masses was 
investigated more thoroughly by Chan, Chattopadhyay and Nath in Ref. \cite{ccn}, where it was noted that the
$\mu$ parameter can be regarded as a measure of fine-tuning, and where the trajectory of constant $\mu$ was found
to form a hyperbolic branch trajectory in the mSUGRA parameter space. In Ref. \cite{fmm}, Feng, Matchev and Moroi
noted the focusing behavior of the $m_{H_u}^2$ renormalization group (RG) trajectory, wherein a variety of GUT scale $m_{H_u}^2$
values would be ``focused'' to a common weak scale value of $m_{H_u}^2$. Since REWSB leads to $\mu^2\sim -m_{H_u}^2$
at the weak scale, the focused solutions gave rise to small $\mu^2$ values. These authors moreover performed a 
sophisticated fine-tuning analysis, and showed that fine-tuning was small in the HB/FP region as long
as $m_{1/2}$ was not too large. It could be seen in Ref. \cite{bb2}, and more fully in \cite{fmw,bbb,balazs}, that the
relic density is indeed low in the HB/FP region, as is typical of mixed higgsino dark matter. Further, in
Ref. \cite{fmw}, it was shown that a variety of direct\cite{nezri,bbbo} and indirect\cite{bo,bbko} 
DM detection rates were large in this region, due
to the large neutralino-nucleon scattering cross sections, and also due to the large neutralino-neutralino
annihilation rates. In addition, collider reaches in the HB/FP region were found for the Fermilab Tevatron\cite{bkt},
the CERN LHC\cite{bcpt,lhcreach,bbbkt} and the International Linear Collider (ILC)\cite{bbktnew}.

All of the above HB/FP analysis were performed in the $m_0\ vs.\ m_{1/2}$ plane of the mSUGRA model. 
However, it was noted in Ref. \cite{bkt} that the exact location of the HB/FP region in the mSUGRA plane
is extremely sensitive to the assumed value of $m_t$. 
In addition, different algorithms for
predicting sparticle masses in the mSUGRA model were found to give very different portrayals of the shape and
location of the HB/FP region even for the {\it same} assumed value of 
$m_t$\cite{kraml}\footnote{Even updated versions of the same
computer codes would give shifted locations of the HB/FP region.}. Finally, it was shown in 
Ref. \cite{nuhm1,nuhm2} that the HB/FP region occurs also in models with non-universal scalar masses,
and that its location can shift depending on the amount and type of non-universality which is assumed. 

In this paper, we propose a more sensible parametrization of the HB/FP region based purely on
weak scale parameters $\mu$ and $U(1)$ gaugino mass $M_1$. Such a parametrization should be
independent of the GUT-to-weak scale evolution algorithm assumed, and also should be only weakly
dependent on the value of $m_t$ or other SUGRA parameters which are assumed. 
In Sec. \ref{sec:fp}, we present our re-parametrization of the HB/FP region, and show
the regions of the weak scale $M_1\ vs.\ \mu$ plane which are allowed by the recent WMAP measurement 
$\Omega_{CDM}h^2=0.113\pm 0.009$ of the relic density of cold dark matter
in the universe. In Sec. \ref{sec:dm}, we show current astrophysical constraints on the HB/FP
region arising from various sources, including an overproduction of ${}^6$Li in the early universe.
We also show the prospects for exploring the HB/FP region via direct and indirect DM detection.
In the HB/FP region, prospects are especially encouraging via Stage 3 direct detection experiments,
detection of DM via neutrino telescopes, and via anti-deuteron searches at experiments such as GAPS.
The latter results depend strongly on the galactic DM  density profile which is assumed.
In Sec. \ref{sec:col}, we show the reach prospects of the CERN LHC and also for the ILC
for SUSY in the WMAP allowed part of the HB/FP region. The CERN LHC 
reach is evaluated in the case of gluino and squark cascade decays, and also for clean trileptons arising
from $\tw_1\tz_2$ production. In both cases, the LHC reach is out to $m_{\tg}\sim 1.7$ TeV 
assuming 100 fb$^{-1}$ of integrated luminosity. The ILC reach is maximal in the $e^+e^-\to \tz_1\tz_2$ 
or $\tz_1\tz_3$ channels, and extends out to $m_{\tg}\sim 1.7$ TeV for a $\sqrt{s}=0.5$ TeV ILC, but to
$m_{\tg}\sim 3.5$ TeV for a $\sqrt{s}=1$ TeV collider.

\section{Unraveling the Focus Point Region}
\label{sec:fp}

The purpose of this section is to introduce the HB/FP parameter space region of the mSUGRA model which produces 
a sufficiently low relic neutralino abundance in the Early Universe at large values of the universal scalar mass $m_0$.
We motivate here why a GUT-scale description of the HB/FP region is ill-suited for phenomenological studies, 
and outline an alternative and complementary weak scale parameterization. 
The latter allows us to describe in detail the cosmologically allowed areas in the more physical 
weak-scale parameter space, and will be used throughout the remainder of this report to study the 
corresponding phenomenology at dark matter search experiments and at colliders.

Radiative electroweak symmetry breaking is the mechanism in which EWSB is triggered by $m^2_{H_u}$ turning 
negative due to its RG evolution. The RGE for $m_{H_u}^2$ reads
\begin{equation}\label{eq:fprge}
\frac{{\rm d}m^2_{H_u}}{{\rm d}t}=\frac{2}{16\pi^2}\left(-\frac{3}{5}g_1^2 M_1^2-3g_2^2M_2^2+\frac{3}{10}g_1^2S+3f_t X_t\right),
\end{equation}
where
\begin{eqnarray}
X_t&=&\left(m^2_{Q_3}+m^2_{U_3}+m^2_{H_u}+A^2_t\right),\\
S&=& m_{H_u}^2-m_{H_d}^2+ Tr\left[ {\bf m}_Q^2-{\bf m}_L^2 
-2{\bf m}_U^2+{\bf m}_D^2+{\bf m}_E^2 \right] ,
\end{eqnarray}
where $S\equiv 0$ at all scales in models with universality.
Among the necessary conditions for the spontaneous breaking of the EWS, the one which determines 
the (absolute) value of the $\mu$ parameter reads (at tree level)
\begin{equation}\label{eq:rewsb}
\mu^2=\frac{m^2_{H_d}-m^2_{H_u}\tan^2\beta}{\tan^2\beta-1}-\frac{m_Z^2}{2}.
\end{equation}
At moderate to large values of $\tan\beta$, as implied by the LEP2 lower bound on the lightest Higgs mass, 
$\mu^2\simeq-m^2_{H_u}$. When the universal mSUGRA GUT scale scalar mass $m_0$ takes values much 
larger than all other soft-breaking masses, the RG evolution of $m^2_{H_u}$ is dominated by the term $X_t$ in Eq.~(\ref{eq:fprge}). 
As $m_0$ increases, cancellations occurring in $X_t$ yield smaller and smaller absolute values for $m^2_{H_u}$. 
Therefore, Eq.~(\ref{eq:rewsb}) leads to the possibility of achieving, in principle, arbitrarily low values for $\mu^2$, until, eventually, $\mu^2<0$, and REWSB can no longer be obtained. 

Within the mSUGRA model, low values of $\mu$ therefore occur in the region of very large universal scalar mass $m_0\gtrsim1$ TeV. This opens up a qualitatively new window in the model parameter space, as far as the cosmological abundance of thermally produced relic neutralinos is concerned. As $\mu$ approaches the low-scale value of the lightest soft-breaking gaugino mass, $M_1$ in the case of mSUGRA, the higgsino component of the LSP increases, leading to an enhancement of efficient LSP pair annihilations into gauge bosons (the latter being largely suppressed in the case of a bino-like LSP). Further, the lightest chargino and the two next-to-lightest neutralinos get closer in mass to the LSP, contributing as well to the suppression of the LSP relic abundance through co-annihilations.

As a result, the particle spectrum of the cosmologically allowed HB/FP region of the mSUGRA model is characterized by (1) a heavy scalar sector, in the multi-TeV range (with the exception of the lightest CP-even Higgs boson), and (2) low values (anywhere below or around 1 TeV) of the $\mu$ parameter. For $\mu\lesssim$ 1 TeV, the requirement of a sufficiently low neutralino relic density forces moreover the relation $\mu\simeq M_1$, the latter being the soft-breaking hypercharge gaugino mass. The weak-scale values of the three soft-breaking gaugino masses, unified to a universal value $m_{1/2}$ at the grand unification (GUT) scale, are given by the usual GUT relations,
\begin{eqnarray}
M_1 & \simeq & 0.44\ m_{1/2}\label{eq:m1}\\
M_2 & = & M_1\ \frac{3}{5\tan^2\theta_W}\\
M_3 & = & M_2\ \frac{\alpha_s(m_Z)\ \sin^2\theta_W}{\alpha_{\rm em}(m_Z)} .
\label{eq:m3}
\end{eqnarray}

Evidently, the decoupling of the supersymmetric sfermion and heavy Higgs sectors implies that the critical parameters entering the phenomenology of the HB/FP region at colliders and at dark matter search experiments are $M_1$ and $\mu$, 
their relative size setting the LSP mass and higgsino fraction. Further, the gluino mass is determined as well by $M_1$, 
through Eqs.~(\ref{eq:m1})-(\ref{eq:m3}). 
The sign of $\mu$ and $\tan\beta$ affect as well, though less critically, the low-energy implications of the setup. 

In the standard GUT-scale parameterization, where one slices the parameter space {\em e.g.} along $(m_0, m_{1/2})$ planes, 
the HB/FP region appears as a narrow line squeezed on the large $m_0$ region adjacent to where no REWSB is attainable. 
The steep and complicated behavior of $\mu$ as a function of $m_0$ makes it difficult to read out the neutralino mass and 
composition in that representation. Further, the HB/FP region is often plagued by numerical stability problems, which will be discussed in detail in Sec.~\ref{sec:numerical}, affecting the phenomenologically crucial low-scale value of the $\mu$ parameter. As a consequence, it is not easy to read out from the standard $(m_0, m_{1/2})$ parameterization most of the phenomenologically relevant information, {\em e.g.} the neutralino mass range and composition compatible with the WMAP relic abundance, and the projected reach of collider and dark matter search experiments.

The purpose of our analysis is therefore to map the GUT-scale representation of mSUGRA in terms of the universal 
scalar and gaugino masses onto the more {\em physical} $(M_1,\mu)$ plane on which the whole HB/FP region 
phenomenology sensitively depends. 
\FIGURE[!t]{\epsfig{file=plot.10.174.eps,width=7.2cm}\quad
\epsfig{file=plot.50.174.eps,width=7.2cm} 
\caption{The projection, onto the ($M_1,\mu$) plane, of some mSUGRA parameter space slices at constant $m_0$, 
for $m_t=174.3$ GeV, $A_0=0$, $\mu>0$, at $\tan\beta=10$ (left panel) and 50 (right panel). 
The blue lines indicate the points featuring the maximal value of $\mu$ at fixed $m_{1/2}$.}
\label{fig:map174}}
\FIGURE[!t]{\epsfig{file=plot.10.178.eps,width=7.2cm}\quad
\epsfig{file=plot.50.178.eps,width=7.2cm} 
\caption{Same as in Fig.~\ref{fig:map174}, but at $m_t=178$ GeV.}
\label{fig:map178}}

From the high-energy scale point of view, $M_1$ depends essentially only on $m_{1/2}$ (see Eq.~(\ref{eq:m1})), 
while $\mu$ is sensitive, in principle, to all mSUGRA GUT input parameters. For given fixed values of $\tan\beta$ and of the trilinear coupling $A_0$, the mSUGRA parameter space spans a limited region only, on the physical $(M_1,\mu)$ plane. 
The $\mu$ parameter features, in fact, a certain maximal value (reached at low or intermediate values of $m_0$) along slices at fixed $m_{1/2}$. Values larger than the mentioned maximum cannot be obtained if REWSB is required. To illustrate this point, we show, in Fig.~\ref{fig:map174}, curves at fixed $m_0$ on the $(M_1,\mu)$ plane, computed with Isajet 7.72~\cite{isajet} at two values of $\tan\beta=10$ and $50$, with an input top mass $m_{\rm t}=174.3$ GeV. 
The iso-$m_0$ curves terminate where the RG code no longer converges to a stable solution, while the blue curves indicate the maximal $\mu$ values. We remark that the low $\mu$ termination point of the curves is in fact only a 
{\em numerical artifact}, and that {\em the entire low $\mu$ region extending to $|\mu | =0$ 
is a physically viable portion of the parameter space, which is not accessible in many codes which rely on 
a GUT-scale parameterization}. 

As expected from Eq.~(\ref{eq:rewsb}), a given value of $\mu$ is obtained, at smaller $\tan\beta$, with larger values of $m_0$, and vice-versa: this reflects the well-known fact that the HB/FP region appears at smaller values of $m_0$ the larger $\tan\beta$ is.

The sensitivity of the HB/FP region parameter space in the ($m_0,\ m_{1/2}$) plane on the value of $m_t$ has been widely reported, 
see {\em e.g.} Ref.~\cite{bkt,Allanach:2003jw,kraml}, and can be readily understood from Eq.~(\ref{eq:fprge}). 
The shape of the weak scale parameter space (the $(M_1,\mu)$ plane) will change as well, but is much less sensitive to the assumed
value of $m_t$.
We show the analogue of Fig.~\ref{fig:map174} but at $m_t=178$ GeV in Fig.~\ref{fig:map178}.

\subsection{Numerical issues in the HB/FP region}\label{sec:numerical}
\FIGURE[!t]{\epsfig{file=suspect_10.eps,width=7.2cm}\quad
\epsfig{file=suspect_50.eps,width=7.2cm} 
\caption{A comparison, along the same parameter space slices as in Fig.~\ref{fig:map174}, between the two RG evolution numerical codes 
Isajet 7.72 \cite{isajet} (red dashed lines) and SUSPECT 2.34 \cite{Djouadi:2002ze} (black solid lines)}
\label{fig:suspect}}


As we outlined above, a first, and critical, numerical issue in the HB/FP region regards the possibility of achieving a stable and convergent solution for low values of the $\mu$ parameter. To investigate this problem, and to verify the consistency of numerical results in the parameter space region of interest, we compared the latest release of Isajet, v.7.72, with the latest release of another RGE evolution package, Suspect, v.2.34 (for a comparison among the results of these and other numerical codes, see also 
Ref.~\cite{Allanach:2002pz,Allanach:2003jw,kraml}).

Fig.~\ref{fig:suspect} illustrates how the two RG evolution codes project iso-$m_0$ lines onto the 
physical parameter space of the focus point region, the $(M_1,\mu)$ plane. 
We adopt for both numerical codes the same input top mass, which we set to $m_t =174.3$ GeV. 
The lines end, as in Fig.~\ref{fig:map174} and \ref{fig:map178}, where no stable solutions are found, 
in the low $\mu$ portion of the plots. 
We see that in both cases, values of $\mu$ smaller than 100-- 200 GeV in the large $m_{1/2}$ region cannot be numerically resolved. 
Further, although both RG codes feature 2-loop RG running of gauge and Yukawa couplings, 
the disagreement is remarkable (particularly in the low $\mu$ region and at low values of $\tan\beta$).

The calculated value of the $\mu$ parameter as a function of $m_0$ along mSUGRA slices at fixed $m_{1/2}$ 
is shown in Fig.~\ref{fig:m0}. While the agreement among the two codes is excellent in the low $m_0$ end of the plots, when approaching the HB/FP region the values of $\mu$ (including the maxima at fixed $m_{1/2}$) significantly differ. 
The largest differences are found at low $m_{1/2}$ and at low $\tan\beta$.  

A variety of numerical issues may cloud the evaluation of the $\mu$ parameter in 
supersymmetric models connecting the GUT scale to the weak scale. For example:
\begin{itemize}
\item
One problem is that the convergence at low values of $\mu$ will depend in part
on the {\em initial guess} for the supersymmetric masses at the very beginning of the iterative RG evolution process. 
The default guess for the supersymmetric masses used in the two codes respectively reads
\begin{eqnarray}
m_{\rm SUSY}= \sqrt{m_0^2+4M_{1/2}^2} &\qquad& {\rm Isajet\ 7.72}\\
m_{\rm SUSY}= 0.5\ \left(m_0+M_{1/2}\right)+m_Z &\qquad&{\rm SUSPECT\ 2.34}
\end{eqnarray}
Clearly, the different trial values used by the two codes can alter the final convergent solution 
and the value of $\mu$ as well, particularly in the highly fine-tuned region where $\mu\rightarrow 0$. 
\item
A further problem occurs in that SUSPECT 2.34 uses two-loop RGEs only for gauge and
Yukawa couplings, but not for soft SUSY breaking terms, while Isajet uses two-loop RGEs throughout its RG treatment.
\item
Another difficulty occurs in that loop corrections must be included in the formulae for the minimization of the scalar
potential of the theory. These loop corrections depend on the full spectrum of supersymmetric particles. 
However, to calculate
the full spectrum, the value of $\mu$ must be known. It is possible in the low $\mu$ region that the tree level value of 
$\mu^2 <0$, while loop corrections will lift $\mu^2 >0$. In this case, a {\it guess} must be made as to what the loop
corrected value of $\mu$ is, just so that a viable spectrum of SUSY particles can be calculated, and used as input, in its turn, 
to the loop corrections. The value of $\mu$ will depend on how this guess is made.  
\item As evident from Eq.~(\ref{eq:fprge}), at very large values of the common scalar mass, 
$\mu$ is extremely sensitive to the top Yukawa coupling $f_t$. The latter is defined as 
\begin{equation}\label{eq:mtop}
f_t=\sqrt{2} \ m_t/v_u,
\end{equation}
where $m_t$ is the running top quark mass in the $\overline{\rm DR}$ scheme. 
The computation of $m_t$ suffers from uncertainties related to (1) the extraction of the $\overline{\rm DR}$ top mass from 
its pole or $\overline{\rm MS}$ values ({\em i.e.} the inclusion of Standard Model threshold corrections) and 
(2) the implementation of the SUSY loop corrections, and the scale choice at which these are implemented.
Similar ambiguities pertain to the evaluation of the bottom and tau Yukawa couplings.
\item The numerical results will depend on the scale choice at which different SUSY particles are integrated out of 
the effective theory. A related issue is the choice of scale at which various SUSY threshold corrections are implemented. 
For instance, SoftSUSY and Spheno assume the MSSM is valid from $M_{GUT}$ all the way to $M_Z$.
When scalars have masses of several TeV, such as in the HB/FP region, this may not be a good assumption.
\end{itemize}

\FIGURE[!t]{\epsfig{file=m0_10.eps,width=7.2cm}\quad
\epsfig{file=m0_50.eps,width=7.2cm} 
\caption{A comparison, at $m_t= 174.3$ GeV and along iso-$m_{1/2}$ slices, 
of the $\mu$ parameter as a function of $m_0$ between the two RG evolution numerical codes 
Isajet 7.72 \cite{isajet} (red dashed lines) and SUSPECT 2.34 \cite{Djouadi:2002ze} (black solid lines), at $\tan\beta=10$ (left) and 50 (right).}
\label{fig:m0}}

As a bottom line, large numerical uncertainties plague the study of the phenomenology of the HB/FP region at the GUT scale, 
depending on a number of assumptions in the details of the RG evolution which lead to significant discrepancies in the determination of $\mu$ from REWSB; on top of this computational ambiguity, a possibly larger uncertainty stems from the input value of the top quark mass: even at the level of accuracy with which $m_t$ will be measured at the LHC the induced variations in the low-scale parameters in the HB/FP region would give rise to completely different phenomenological scenarios at the same GUT-scale input values \cite{kraml}. Lastly, currently available numerical codes do not fully access the phenomenologically interesting 
mSUGRA region at $\mu\ll M_1$, which can thus be only explored resorting to a low-scale parameterization.

\subsection{Outline of the low-energy parameterization}\label{sec:parameterization}

We described above how the procedure of outlining a systematic mapping between the phenomenologically relevant low-energy parameter space and the customary mSUGRA GUT-scale input setting faces a number of intrinsic issues. Moreover, those issues evidently blur the possibility of a bottom-up reconstruction of mSUGRA high-energy input parameters. Our main point here is therefore that {\em the collider and dark matter phenomenology of the focus point region is better studied through a convenient two-parameters low-energy description, which is complementary to the usual GUT-scale parameterization, and which thoroughly reproduces all phenomenological implications of the ``mother theory'' at the high-energy scale}. This section is henceforth devoted to the construction of such a low-scale parametrization, of which we will then make use in the remainder of this report.

We argued above that the physical parameter space of the HB/FP region is given by the set $(M_1,\mu,\tan\beta)$. At a given value of $\tan\beta$, $M_1$ (and therefore of $m_{1/2}$ at the GUT scale) and of the input top quark mass 
$m_t$, REWSB bounds the range of $\mu$ from above, through the condition given in Eq.~(\ref{eq:rewsb}). As pointed out in the previous section, the precise boundaries of the region allowed by REWSB will also depend on numerical subtleties. However, we will show below that the REWSB boundary at large $\mu\gg M_1$ lies outside the cosmologically relevant HB/FP region. As a result, these uncertainties will not affect the discussion of the HB/FP region phenomenology, since that portion of the parameter space will be disregarded after the neutralino relic density analysis. The other soft breaking gaugino masses $M_2$ and $M_3$ are given, as functions of $M_1$, by the GUT relations specified in 
Eqs.~(\ref{eq:m1})-(\ref{eq:m3}).

Since the heavy scalar sector is largely decoupled from the low-energy phenomenology in the HB/FP region, the details of the sfermions and heavy Higgses spectrum play a very marginal role. In this respect, we resort in most of our plots, to setting, for simplicity, all sfermions and heavy Higgs masses to a common value $m_{\tilde f}=5 $ TeV, much larger than all other relevant weak-scale parameters, and close to the average value of the masses obtained by a full RGE treatment. Further, we set all trilinear couplings to zero.

The value of the lightest Higgs mass $m_h$ enters, instead, quite critically in a few quantities, in particular those related to the neutralino scattering off matter: for instance, the neutralino-quark spin independent scattering cross section $\sigma^{SI}_{\tz_1 q}$, in the limit of large scalar masses, is dominated by $t$-channel light Higgs exchanges, and scales as $\sigma^{SI}_{\tz_1 q}\propto m_h^{-4}$. In mSUGRA, the value of the lightest Higgs mass depends, in principle, on all input SUSY parameters. In the HB/FP region, at large universal scalar masses, we find a critical dependence on $m_{1/2}$, while the GUT-scale values of $m_0$ and $A_0$ are far less important. Consistently with the assumption of a common low-scale scalar mass $m_{\tilde f}$, we therefore expressed $m_h$ as a function of the parameters $M_1$ and $\tan\beta$, according to a phenomenological functional dependence of the form
\begin{equation}
m_h=m^0_h\ \left(\frac{M_1}{M_1^0}\right)^\gamma.\label{eq:mh}
\end{equation} 
The formula was then fitted against mSUGRA points randomly generated at large values of $m_0$. 
We find that $m^0_h$ and $m_M^0$ do not critically depend on $\tan\beta$, and the best fit values read $m^0_h=115.5$ GeV and $M_1^0=100$ GeV, respectively. 
The exponent $\gamma$ depends instead more sensitively on $\tan\beta$: for instance, 
we find $\gamma\simeq 0.03$ for $\tan\beta=10$, and $\gamma\simeq 0.0335$ for $\tan\beta=50$. 
We estimate the typical accuracy of Eq.~(\ref{eq:mh}) in reproducing the correct value of $m_h$ to be less than 1 GeV for $M_1\gtrsim 300$ GeV, 
and within 2 GeV for values of $M_1$ smaller than 300 GeV. 
The relative error induced in $\sigma^{SI}_{\tz_1 q}$ is therefore expected to be of $\mathcal{O}(1\%)$.
 
\subsection{Neutralino relic abundance}\label{sec:relicabundance}

\FIGURE[!t]{\epsfig{file=50_wide.eps,width=8.2cm}\quad
\epsfig{file=relic.eps,width=6cm} 
\caption{(Left): The allowed region, on the ($M_1,\mu$) plane, at $\tan\beta=50$. 
The area shaded in red features a lightest chargino mass below the LEP2 limit \cite{lep2}, i.e. $m_{\chi^+}<103.5$ GeV. 
In the gray shaded area the value of $\mu$ exceeds the maximal one allowed by radiative electroweak symmetry breaking (see Fig.~\ref{fig:map174}). The yellow shaded area is cosmologically excluded since the thermally produced neutralino relic density exceeds the 2-$\sigma$ upper bound on the cold dark matter abundance. Finally, points in the green region feature a neutralino thermal relic abundance compatible with the upper and lower limits on the cold dark matter content of the Universe. We also indicate the location of the coannihilation strip in the ($M_1,\mu$) plane. 
The solid black lines indicate those points giving exactly the central value $\Omega_{\tz_1} h^2=0.11$ in the HB/FP region. 
(Right): The neutralino thermal relic abundance as a function of $\mu$ at fixed $M_1=250$ GeV (solid black line), 1000 GeV (red dashed line) and 1500 GeV (blue dot-dashed line). The green strip indicates the WMAP favored range for the cold dark matter abundance.}
\label{fig:paramspace}}

Large values of the sfermion masses in the HB/FP region help alleviate a number of well-known phenomenological difficulties of supersymmetric extensions of the standard model, ranging from the SUSY flavor and $CP$ problems, to dimension-five proton decay operators which often appear in SUSY-GUT embeddings. In the present framework, where we neglect $CP$ violating phases, and assume a minimal flavor structure, the SUSY contributions to rare processes, {\em e.g.} the branching ratios $b\rightarrow s\gamma$, $B_s\rightarrow\mu^+\mu^-$, or to the muon anomalous magnetic moment, are very suppressed, being mediated by sfermions loops, and do not constrain the HB/FP region parameter space. Further, limits from direct sfermion searches at colliders are always very distant from the sfermion mass scales of the HB/FP region.

The phenomenological constraints which apply to the HB/FP region of mSUGRA are henceforth limited to the LEP2 searches for the lightest chargino, and to gluino searches at the Tevatron. In the present setting with universal gaugino masses, the latter bound is however much less constraining than that stemming from chargino searches. We use here the mass limit $m_{\tw_1}>103.5$ GeV \cite{lep2}, although in the pure-higgsino region 
($\mu\ll M_1$) the LEP2 bound is weakened by the quasi-degeneracy between the lightest chargino and the LSP (namely, $m_{\tw_1}>92.4$ GeV \cite{lep2}). The ($M_1,\mu$) plane is further constrained, at a given value of $\tan\beta$, by the REWSB conditions which limit the range of $\mu$ from above. In the left-over portion of parameter space we compute the thermal LSP relic abundance $\Omega_{\tz_1}h^2$ with the {\tt DarkSUSY} package~\cite{Gondolo:2004sc}. We rule out models giving $\Omega_{\tz_1}h^2>0.13$, according to the 2-$\sigma$ upper bound on the cold dark matter abundance derived by the WMAP collaboration \cite{Spergel:2003cb}

In Fig.~\ref{fig:paramspace}, left, we plot the parameter space of the HB/FP region using the scheme outlined in Sec.~\ref{sec:parameterization}, at $\tan\beta=50$ and positive $\mu$. In this plot only, 
the scalar masses are determined by the value of $m_0$ needed to give the appropriate $\mu$ value 
as in Fig. \ref{fig:m0}{\it b}.The region shaded in red is excluded by LEP2 chargino searches, while in the gray shaded area the value of $\mu$ exceeds the maximal value compatible with REWSB at $\tan\beta=50$ and $\mu>0$ (see Fig.~\ref{fig:map174}, right). The parameter space portion shaded in yellow indicates where the upper bound on $\Omega_{\tz_1}h^2$ is violated. The viable parameter space of the theory is thus restricted to the green band (giving $\Omega_{\tz_1}h^2$ {\em within} the 2-$\sigma$ WMAP range) and to the region in white ($\Omega_{\tz_1}h^2$ {\em below} the 2-$\sigma$ WMAP range). 
In the low-relic density white region we suppose that some mechanism in the Early Universe has enhanced the final relic density of neutralinos ({\em e.g.} through non-thermal neutralino production \cite{non-therm}, or through a modified cosmic expansion at the time of neutralino freeze-out, as it might be the case in scenarios with quintessence \cite{quint}, with a Brans-Dicke-Jordan modified theory of gravity \cite{shear,Catena:2004ba} or with an anisotropic primordial expansion of the Universe \cite{shear,Profumo:2004ex}). We therefore work under the assumption that all CDM 
is composed of relic neutralinos.
\FIGURE[!t]{\epsfig{file=neut_mass.eps,width=7.2cm}\quad
\epsfig{file=neut_comp.eps,width=7.2cm} 
\caption{The iso-level curves for the lightest neutralino mass (left) and composition (in terms of the higgsino fraction $Z_h\equiv N_{13}^2+N_{14}^2$, right), on the parameter space outlined in Fig.~\ref{fig:paramspace}, at $\tan\beta=50$ and $\mu>0$.}
\label{fig:neut}}

For completeness, we also indicate in Fig.~\ref{fig:paramspace} the location of the stau coannihilation strip in this low-energy representation, squeezed along the line of maximal $\mu$ values. Had we chosen to consider negative values for $\mu$, a rapid heavy Higgs $s$-channel exchange-mediated annihilation funnel would also have opened. The latter would lie at rather large values of $\mu$ (slightly below the coannihilation strip, since the corresponding $m_0$ would have been larger in the funnel than in the coannihilation region, and therefore the corresponding $\mu$ range would be slightly lower (see Fig.~\ref{fig:m0}), possibly extending to even larger values of $M_1$. In this respect, the $(M_1,\mu)$ plane is 
{\em not} a good representation of the coannihilation and funnel regions of mSUGRA, since it trades a physically relevant parameter for those regions ($m_0$) for a parameter which is instead not critical for $\Omega_{\tz_1}h^2$, {\em i.e.} $\mu$. We therefore stress a {\em strong complementary role} of the GUT-scale and of the present low-scale parameterizations for the phenomenological study of the mSUGRA model.

The shape of the parameter space region giving exactly the central value of the CDM density determined by WMAP is indicated with a black line approximately lying in the center of the green-shaded 2-$\sigma$ area. The shape of that line provides a non-trivial information on the mechanisms responsible for a WMAP-compatible neutralino thermal relic abundance in the HB/FP region. In the large $m_{1/2}$ (large $M_1$) limit, the neutralino relic abundance corresponds to that of a pure higgsino-like LSP, and is fitted, with high precision, by the formula
\begin{equation}
\Omega_{\tz_1}h^2\simeq 0.1\left(\frac{\mu}{{\rm 1\ TeV}}\right)^{1.906} .
\end{equation}
This formula entails the important result that {\em the maximal neutralino mass compatible at 95\% C.L. with the WMAP upper limit on the CDM abundance in the minimal supergravity model corresponds to} 1150 GeV (We recall that in the funnel and coannihilation regions $m_{\tz_1}$ is always found to be less than 900 GeV, see {\em e.g.} \cite{Ellis:2003cw}). On the other hand, the central WMAP CDM abundance value here is attained at $m_{\tz_1}\simeq 1050$ GeV. 

The central part of the curve lies along the $M_1\simeq \mu$ edge; 
in this region the interplay of coannihilation processes and of a mixed bino-higgsino LSP cooperates to fulfill the condition $\Omega_{\tz_1}h^2\simeq\Omega_{\rm CDM}h^2$. In the $(M_1,\mu)$ plane the 2-$\sigma$ range of $\mu$ shrinks: this fact depends on the transition from the pure higgsino to the pure bino regime. We illustrate explicitly this phenomenon in the right panel of fig.~\ref{fig:paramspace}, where we show $\Omega_{\tz_1}h^2$ as a function of $\mu$ along slices at fixed $M_1$. The mentioned transition, which we indicate as ``coannihilation regime'' in the figure, starts at 
$\mu\simeq M_1$ and ends when $\mu\gg M_1$; 
in this regime, the relic abundance is clearly a very steep function of $\mu$.

As $M_1$ is further decreased, the neutralino mass compatible with the WMAP CDM relic abundance decreases, while the bino fraction and the mass splitting between the LSP and the lightest chargino increase. This is illustrated in Fig.~\ref{fig:neut}, which also shows the iso-neutralino mass contours and the iso-higgsino fraction levels on the $(M_1,\mu)$ plane, which will be the object of our phenomenological analysis in the remainder of this report.

In particular, the right panel of Fig.~\ref{fig:neut} illustrates that for $m_{1/2}\lesssim2.5$ TeV the lightest neutralino in the HB/FP region is always a very mixed bino-higgsino particle, with higgsino fractions $Z_h\equiv N_{13}^2+N_{14}^2$ between 0.1 and 0.9. At larger $m_{1/2}$, instead, $Z_h>0.9$, and the ``pure higgsino regime'' ($Z_h>0.99$) is reached at $m_{1/2}\gtrsim3$ TeV.

Finally, let us point out the threshold effects, at $M_1\simeq 200$ GeV and $M_1\simeq 100$ GeV, 
respectively, corresponding to $m_{\tz_1}\simeq m_t$ and $m_{\tz_1}\simeq m_{W,Z}$. 
When one of the $t\bar t$, $WW$, $ZZ$ final channels closes up, the resulting suppression of the lightest neutralino pair annihilation cross section must be compensated with an increased higgsino fraction and a reduced mass splitting between the LSP and its coannihilation partners, resulting in the above mentioned bumps at the corresponding thresholds.

As a concluding remark, we stress that the picture we outlined above is {\em very mildly dependent on the particular value of} $\tan\beta$ {\em we picked}, and on the sign of $\mu$ as well. We explicitly worked out the WMAP allowed parameter space for lower $\tan\beta$, and the emerging picture is almost indistinguishable from what we show here, providing evidence for a kind of {\em universality} in the HB/FP region phenomenology on the 
$(M_1,\mu)$ plane. In particular, our conclusions on the maximal LSP mass in mSUGRA are not significantly affected.

\section{Dark Matter Phenomenology}
\label{sec:dm}

In the present section we study the implications of SUSY models belonging to the HB/FP region parameter space outlined in 
Sec.~\ref{sec:relicabundance} for dark matter searches. Following previous analysis~\cite{pierohalos,Edsjo:2004pf,Profumo:2004ty}, we make use of two extreme halo profiles, a cored halo model (the {\em Burkert profile}, see \cite{elzant,burkert,salucci}) and an Adiabatically contracted version \cite{blumental} of the cuspy halo model proposed in Ref.~\cite{n03} (which we dub {\em Adiabatically Contracted N03 profile}). We claim that those two instances are indicative of the range of variations in dark matter detection rates induced by different consistent models of the dark matter distribution in the halo, respectively, giving minimal (Burkert) and maximal (Adiab.Contr.N03) rates. 
We refer the reader to Ref.~\cite{pierohalos} for details. 

\FIGURE[!t]{\epsfig{file=50.BUR.cur.eps,width=7.2cm}\quad
\epsfig{file=50.N03.cur.eps,width=7.2cm} 
\caption{Current constraints on the ($M_1,\mu$) plane from dark matter physics, at $\tan\beta=50$, $\mu>0$, for the Burkert Halo Model (left) and for the Adiabatically contracted N-03 Halo Model (right). The gray shaded area is excluded by the LEP2 searches for the lightest chargino and for an excessive neutralino thermal relic abundance (see Fig.~\ref{fig:paramspace}), while the green shaded region is the 2-$\sigma$ WMAP-favored region from the neutralino relic abundance. The area shaded in dark blue is ruled out by current data on the antiprotons flux, while the area shaded in brown by the primordial $^6$Li abundance bound. In the plot to the right, the cyan region is not consistent with positrons flux data, and the region shaded in black gives an excessive gamma-ray flux in the energy range of the EGRET experiment.}
\label{fig:dm.cur.50}}

Sec.~\ref{sec:current} is devoted to a parallel analysis, for the two halo models, of the current constraints coming from antimatter and gamma ray fluxes. We also impose the constraint coming from the overproduction of ${}^6$Li in the Early Universe \cite{Jedamzik:2004ip,Masiero:2004ft}. Sec.~\ref{sec:future} is instead devoted to prospects for dark matter detection in the HB/FP region at future experiments, 
while in Sec.~\ref{sec:antimatter}, we study the spectral features at future space-borne antimatter experiments of models in the HB/FP region giving a thermal relic density of neutralinos consistent with the WMAP CDM abundance.

As mentioned in Sec.~\ref{sec:relicabundance}, we assume here that models with a low thermal relic abundance, 
in the $\mu\ll M_1$ region, are responsible for all the CDM in the Universe, in virtue of the above mentioned  
mechanisms of relic density enhancement (see {\it e.g.} Ref.~\cite{Profumo:2004ex}).

\subsection{Overview of current constraints}\label{sec:current}

Current dark matter detection results include the recently released direct detection exclusion limits delivered by the CDMS collaboration \cite{Akerib:2004fq}, the SuperKamiokande upper limit on the muon flux from neutralino pair annihilation in the core of the Sun \cite{Habig:2001ei}, the antimatter fluxes as measured by balloon-borne experiments \cite{antimatterref} and the EGRET data \cite{egret} on the gamma ray flux from the Galactic Center.
Besides direct and indirect dark matter searches, neutralinos are also constrained by the synthesis of ${}^6$Li in post-freeze-out neutralino annihilations, as recently shown in Ref.~\cite{Jedamzik:2004ip}. This constraint can be rephrased as a constraint on the neutralino pair-annihilation cross section \cite{Jedamzik:2004ip}, and, contrary to dark matter search results, is independent of the halo model under consideration.

In Fig.~\ref{fig:dm.cur.50}, we depict the above mentioned current exclusion limits in the ($M_1,\mu$) plane. 
The region shaded in gray corresponds to the parameter space excluded either by LEP2 searches for charginos, 
or by the lack of REWSB solutions, or by the overproduction of neutralinos in the Early Universe. The green-shaded area corresponds to the models giving $\Omega_{\tz_1}h^2\simeq\Omega_{\rm CDM}h^2$ at the 2-$\sigma$ level. The left panel corresponds to the Burkert profile, while the right panel to the Adiab.Contr.N03 profile.

The ${}^6$Li constraint is violated on the brown-shaded area, extending in the pure higgsino limit in the mass range $90\ {\rm GeV}\lesssim m_{\tz_1}\lesssim 150\ {\rm GeV}$, where the neutralino pair annihilation cross section $\langle\sigma v\rangle_0$ is maximal.

Neutralino induced primary antimatter fluxes, derived according to the approach described in detail in Ref.~\cite{Profumo:2004ty}, are not consistent, at 95\% C.L., with the measured total (signal plus background) antimatter flux of antiprotons and positrons \cite{antimatterref} in the regions shaded in dark and light blue, respectively. We notice that for both halo models, the antiproton flux constraint is stronger than the ${}^6$Li bound; the positron flux is instead in excess to the measured one only for the cuspy halo model, at $m_{\tz_1}<170$ GeV.

The EGRET data give an energy-dependent upper bound on the gamma ray flux from the galactic center, under the conservative hypothesis that the neutralino pair-annihilation induced signal dominates over a negligible background. The gamma ray flux depends critically on the inner structure of the halo profile, and is therefore greatly enhanced for a cuspy profile, while it is suppressed for a cored halo model. In fact, the EGRET data do not give any constraint if one assumes the Burkert profile, while they rule out pure higgsinos as heavy as 350 GeV with the cuspy Adiab.Contr.N03 profile.

As a final remark, we point out that neither direct detection searches nor current data on the muon flux from the Sun give any constraint on the HB/FP region parameter space.

\subsection{Future dark matter searches}\label{sec:future}

\FIGURE[!t]{
\epsfig{file=50.BUR.fut.eps,width=14cm}
\caption{Reach contours of future dark matter search facilities. The region shaded in gray is excluded in the light of the results shown in 
Fig.~\ref{fig:paramspace} and \ref{fig:dm.cur.50}. The value of $\tan\beta$ is set to 50, $\mu>0$, and the cored Burkert Halo Model is assumed.}
\label{fig:dm.BUR.50}}
\FIGURE[!t]{
\epsfig{file=50.N03.fut.eps,width=14cm} 
\caption{Same as in Fig.~\ref{fig:dm.BUR.50}, but for the Adiabatically contracted N-03 Halo Model}
\label{fig:dm.N03.50}}

We outline in Fig.~\ref{fig:dm.BUR.50} (for the Burkert profile) and \ref{fig:dm.N03.50} (for the Adiab.Contr.N03 profile) the sensitivity reach of a number of future direct and indirect dark matter detection experiments. Models in the area enclosed within each contour give a signal which is larger than the expected sensitivity of the corresponding search facility. 

``Stage-2'' detectors refer to experiments like CDMS2~\cite{cdms2}, Edelweiss2~\cite{edelweiss}, CRESST2~\cite{cresst}, ZEPLIN2~\cite{zeplin}, which will be operative in the near future (the reference sensitivity curve we take here is that corresponding to the CDMS2 experiment \cite{cdms2}). We indicate as ``Stage-3'' detectors ton-size experiments like XENON~\cite{xenon}, Genius~\cite{genius}, ZEPLIN4~\cite{zeplin4} and WARP~\cite{warp} (the reference sensitivity curve is here chosen to be that of XENON~\cite{xenon}).

The reach of Stage-2 detectors is found to be limited to a tiny region at very low masses, largely already ruled out by the chargino mass limit. A neutralino in the HB/FP region is therefore not likely to be detected at Stage 2 direct detection experiments. 
On the other hand, ton-sized detectors look very promising, their sensitivity extending, quite independently of the halo model, and largely independently of $\tan\beta$, over the whole region compatible at 2-$\sigma$ with the WMAP CDM neutralino thermal relic abundance, 
up to $m_{1/2}\lesssim3$ TeV. At larger values of the common gaugino mass, the higgsino fraction becomes exceedingly large (see Fig.~\ref{fig:neut}, right), and, since $\sigma^{SI}_{\tz_1 P}\propto N_h^2(1-N_h)^2$, the resulting direct detection rates are suppressed.

The flux of muons from the Sun is also extremely sensitive to the degree of gaugino-higgsino mixing. The equilibrium between the capture rate inside the Sun and the neutralino pair annihilation, basic in order to produce a sizable neutrino flux out of the Sun, is reached provided the neutralino features (1) a sufficiently large pair annihilation cross section and (2) a large enough spin-dependent neutralino-proton scattering cross section. These two conditions explain the shape of the IceCube reach contours \cite{icecube}, which extend along the largest $\langle\sigma v\rangle_0$ area at low $\mu$ as far as the gaugino fraction is still non-negligible, and in the maximal mixing region. The latter region largely overlaps the WMAP favored green-shaded region. Remarkably enough, neutralinos producing the WMAP required amount of relics to make up the CDM in the Universe tend to maximize, in the HB/FP region, the detection rate at neutrino telescopes! We find that, quite independently of the halo model, the IceCube reach along the cosmologically favored strip extends up to $m_{1/2}\lesssim 1.5$ TeV, corresponding to a neutralino mass between 600 and 700 GeV.

The dependence of the antimatter flux on the halo model \cite{Profumo:2004ty} is found, instead, to be indeed critical. For antiprotons and positrons, the largest fluxes correspond to the pure higgsino region, partially extending into the large bino-higgsino mixing region. The shape of the reach contour for the Pamela experiment (computed following the approach outlined in Ref.~\cite{Profumo:2004ty}) reflects the role of the top threshold (below which a much larger higgsino fraction is needed) and extends into the light neutralino mass region until the second critical threshold ($m_{\tz_1}<m_W$) is reached. Assuming a cored profile, models giving the WMAP relic neutralino abundance do not yield large enough antiprotons and/or positrons fluxes, while with a cuspy profile the reach along the WMAP strip in the antiproton channel extends up to neutralino masses as large as 400 GeV.

Low energy antideuterons have been shown to provide a clean indication of new physics, since the background flux in the 0.1-- 1 GeV 
antideuteron kinetic energy interval is extremely suppressed \cite{dbar}. We assessed the antideuteron flux for the AMS-02 experiment \cite{ams}, which will be sensitive to a flux in the energy band $0.1-2.7$ GeV at the level of $4.8\times 10^{-8}\ {\rm m}^{-2}\ {\rm sr}^{-1}\ {\rm GeV}^{-1}\ {\rm sec}^{-1}$, and for a GAPS experiment \cite{GAPSproposal} placed on a high latitude mission satellite, sensitive to antideuterons in the energy band $0.1-0.4$ GeV at the level of $2.6\times 10^{-9}\ {\rm m}^{-2}\ {\rm sr}^{-1}\ {\rm GeV}^{-1}\ {\rm sec}^{-1}$. We find that {\em antideuteron fluxes accessible to AMS-02 are in general excluded by current bounds on the antiprotons flux, independently of the assumed halo profile}. The GAPS sensitivity extends instead well beyond the antiproton search reach of Pamela in the WMAP favored region, with maximal accessible neutralino masses as large as 400 GeV in the case of a cored profile and of even 700 GeV in the case of a cuspy profile. The antideuteron flux is also greatly sensitive to the gaugino-higgsino mixing, thus, as in the case of the muon flux from the Sun, the maximal reach is gained exactly along the WMAP 2-$\sigma$ region.

The parameter space reach of the GLAST experiment \cite{glast}, finally, largely depends on the details of the inner structure of the halo model under consideration. As can be deduced comparing Fig.~\ref{fig:dm.BUR.50} and \ref{fig:dm.N03.50}, the resulting GLAST reach can be either the 
worst or the best among all direct and indirect detection channels, depending on the amount of dark matter in the center of the Galaxy: for the two halo models under consideration, for instance, this induces a variation of more than four orders of magnitude \cite{nuhm2}! Little can therefore be said about the sensitivity reach of future gamma rays experiments without strong astrophysical assumptions (for recent related 
studies see \cite{gammas,bo,bbko}).

\subsection{Spectral features at space-borne antimatter searches}\label{sec:antimatter}
\FIGURE[!t]{
\epsfig{file=pbep.eps,width=14cm} 
\caption{The signal-over-background for antiprotons (left) and positrons (right), as functions of the antimatter particles' kinetic energy, for four values of the neutralino mass along the WMAP-relic-abundance line at $\tan\beta=50$, $\mu>0$. See the text for details on the background and signal computation.}
\label{fig:pbep}}

In the preceding section we assessed the reach of future space-based antimatter experiments by means of a statistical treatment which evaluates the possibility of disentangling a pure-secondary antimatter flux from the presence of a statistically non-negligible primary component. In case such a signal is detected at Pamela or AMS-02, we will be given the opportunity of studying in some detail the spectral features of the primary component, depending on the relative signal-to-background ($S/B$). In particular, the future wealth of data on antimatter fluxes will greatly reduce the uncertainties in the background determination to an unprecedented level of accuracy. For this reason, we investigate in this section the $S/B$ yields of models in the HB/FP region, concentrating on the parameter space slice giving $\Omega_{\tz_1}h^2$ equal to the central value of the WMAP-inferred CDM abundance $\Omega_{\rm CDM}h^2$.

We show in Fig.~\ref{fig:pbep} the $S/B$ as a function of the antiparticle's kinetic energy for antiprotons (left) and positrons (right), at four different neutralino masses. The antiprotons $S/B$ features in all models a clean peak at $T_{\bar p}\sim0.1m_{\tz_1}$, generated by hard decay modes, particularly from gauge boson decays, plus a low energy tail, mainly fueled by products of $b\bar b$ final state processes. Positrons show a more complex $S/B$, featuring a series of peaks, corresponding to quark jets yielding $\pi^+\rightarrow e^+$ and positrons from $\tau^+$ decays at low energies, and to positrons generated by gauge boson decays at higher energies. In particular, prompt $W$ and $Z$ decays motivate the bump at $T_{e^+}\simeq m_{\tz_1}/2$, which is however very suppressed, and hard to recognize, particularly beyond the $m_{\tz_1}=m_t$ threshold, which tends to soften the positron spectrum.

The location of the maximal ($S/B$) ratios is studied in Fig.~\ref{fig:tmax}, where we plot on the right axis the corresponding antiparticle's kinetic energy, and on the left axis the actual $S/B$ value at the maximum. In both cases, the $m_t$ threshold 
(corresponding to $M_1\simeq 200$ GeV in the plot) is clearly visible. The maxima, in that low neutralino mass end, are located around 10 GeV, for both positrons and antiprotons. For $m_{\tz_1}>m_t$, the location of the antiproton's maximal $S/B$ approximately 
linearly tracks $m_{\tz_1}$, while for positrons the maxima are positioned around 10 GeV, until eventually the $S/B$ corresponding to the $T_{e^+}\simeq m_{\tz_1}/2$ prompt positron production dominates.
\FIGURE[!t]{\epsfig{file=tmaxpb.eps,width=7.3cm}\quad
\epsfig{file=tmaxep.eps,width=7.3cm} 
\caption{On the right y-axis of the two panels we plot the antiparticles' kinetic energy $T_{\rm max}$ at which the maximal $(S/B)$ is attained, in the case of antiprotons (left) and positrons (right). The left y-axis indicate the actual maximal $(S/B)$ at the corresponding $T_{\rm max}$. In the case of positrons (right panel), we also plot the $(S/B)$ at $T_{e^+}=m_{\tz_1}/2$.}
\label{fig:tmax}}

In order to understand to which extent the above analyzed spectral features are specific to the HB/FP region (or to any mixed higgsino-bino LSP scenario), we carried out a comparison of ($S/B$) in different LSP scenarios, namely that of a pure bino (mainly annihilating into $b\bar b$ and $\tau^+\tau^-$ pairs), as in the coannihilation and funnel region of mSUGRA, and that of a pure wino LSP (mainly annihilating into $W^+W^-$ pairs), as in the minimal anomaly mediated supersymmetric breaking model. The case of the bino is characterized by a much larger antimatter yield at low energies, which, for instance, smooths out the maximum in the antiproton $S/B$ pointed out above, and which shifts the positron's maximal $S/B$ towards much lower energies. This LSP scenario is thus virtually distinguishable from the HB/FP region LSP scenario on the basis of a correlated $S/B$ analysis. The case of a wino-like LSP is instead more subtle. The antiproton $S/B$ has a more depressed low-energy tail (which can be however hard to disentangle due to uncertainties in the low-energy antiproton background computation), and the location of the maximal $S/B$ as a function of the mass is at larger $T_{\bar p}/m_{\tz_1}$ values. On the other hand, a possible handle is provided by a much stronger peak in the positron spectrum corresponding to $T_{e^+}\simeq m_{\tz_1}/2$.

The occurrence of a maximum in the antiproton's ($S/B$) correlated with the neutralino mass can evidently be used as an indirect indication of the neutralino mass scale, keeping in mind the above mentioned caveat on a possible entanglement of the mixed-higgsino-bino and wino LSP scenarios. Vice versa, should collider experiments or other dark matter searches point at a certain neutralino mass, the analysis of the antiproton's $S/B$ can be used as a cross-check to understand the nature of the dark matter particle. Positron fluxes appear, instead, less promising for the same task, the absolute values of the $S/B$ being moreover much smaller than those of antiprotons. The exciting perspective of a correlation between the antiproton's 
and the positron's spectral features, which could point to a specific LSP dark matter scenario, therefore also appear rather problematic.


\section{HB/FP region at LHC and ILC}
\label{sec:col}

\subsection{HB/FP region at the CERN LHC}
\FIGURE[!t]{
\epsfig{file=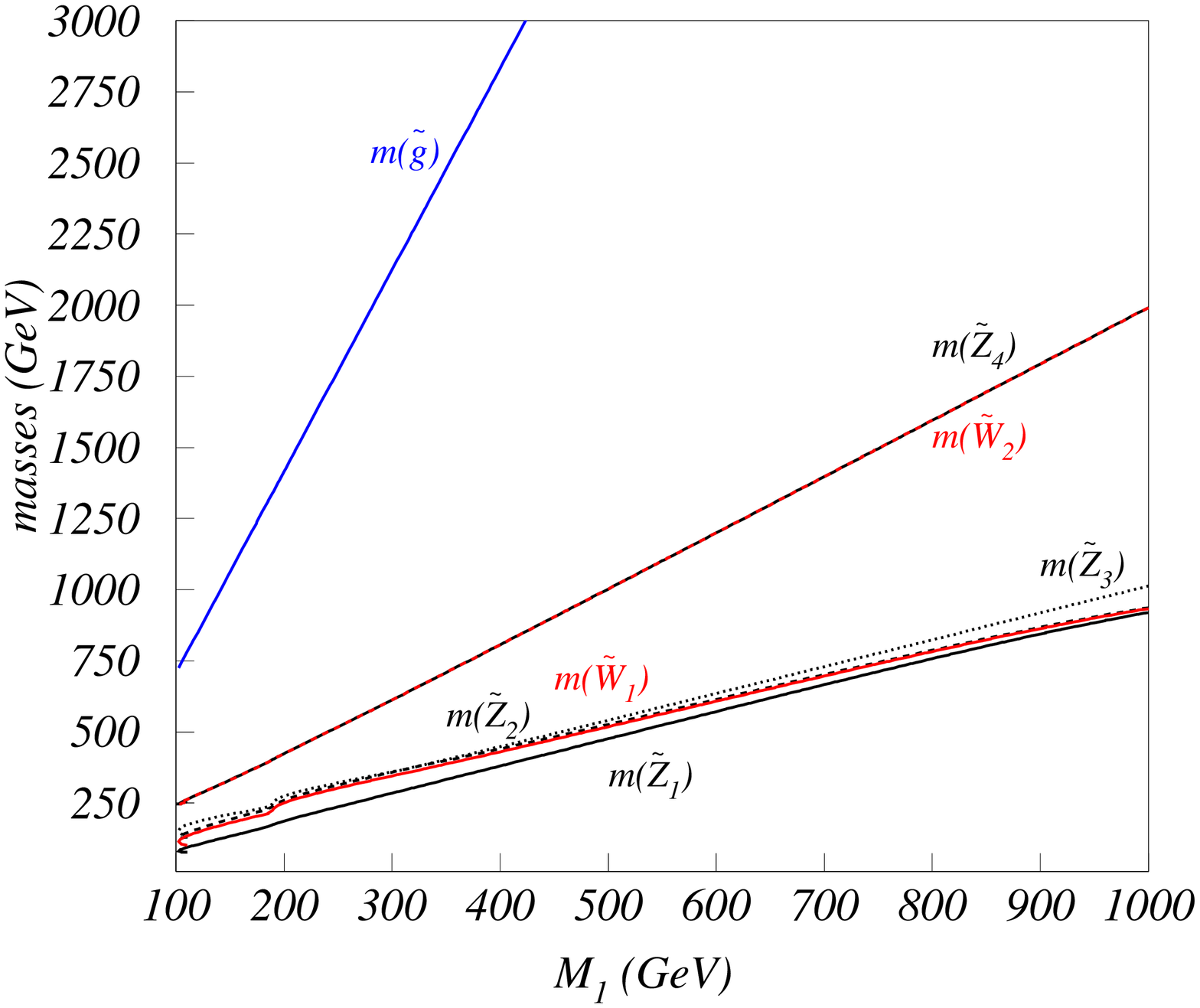,width=7.7cm}\quad
\epsfig{file=split_hfrac.eps,width=6cm}
\caption{mSUGRA with $\tan\beta=50$, $\mu>0$. 
(Left): The mass 
spectrum in the neutralino-chargino sector, 
along the same parameter space line. (Right): The higgsino fraction, 
as a function of $M_1$, along the line at 
$\Omega_{\tz_1} h^2=0.11$ (red dashed line), and the absolute mass splitting 
between the lightest chargino and the lightest neutralino along the 
line at $\Omega_{\tz_1} h^2=0.11$ (black solid line).}
\label{fig:splitting}}

The reach of the LHC in mSUGRA's ($m_0,\ m_{1/2}$) plane was described in 
Ref.~\cite{bbbkt} (see \cite{lhcreach} for related earlier work).
The search strategy was based on the 
detection of gluino and squark cascade decay products, 
namely multiple jets and/or leptons and 
large transverse missing energy. Since sfermion masses in the HB/FP region lie in the multi-TeV range, LHC 
will not produce squarks and sleptons at detectable levels. 
Gluinos can be relatively light ($m_{\tilde g}\lesssim 2$ TeV, 
left frame of Fig.~\ref{fig:splitting}) only in the low $M_1\alt 250$ GeV 
(or, equivalently, low $\mu$) part of the HB/FP region along the line 
$\Omega_{\tz_1} h^2=0.11$. It was found in \cite{bbbkt} that in the HB/FP 
region of mSUGRA, gluino masses of up to $\sim 1.8$ TeV could be probed. 
Therefore one would not expect the method, based 
on the production of heavy strongly interacting superpartners, to work very 
well in the HB/FP region for larger $M_1$ values. We note that 
recent work by Mercadante {\it et al.}\cite{Mercadante:2005vx} 
employed $b$-tagging in their analysis 
and have achieved an up to 20\% extension of the LHC reach in the HB/FP region.

However, as can be seen from the left frame of Fig.~\ref{fig:splitting}, 
charginos and neutralino (collectively dubbed as -inos) 
can still be relatively light in the HB/FP region. 
Even more encouraging is the 
fact that the mass splitting between the lightest chargino and lightest 
neutralino stays $>20$ GeV all the way up to $M_1\sim 850$ GeV and the mass 
gap never exceeds the Z boson mass (right frame of Fig.~\ref{fig:splitting}). 
This means that 3-body decays 
$\tilde W_1\rightarrow\tilde Z_1f\bar f$ and $\tz_2\rightarrow \tz_1 f\bar{f}$ 
are open, and one expects 
multiple leptons in the final 
state. We will resort to the leptonic signals when exploring -ino
production at the LHC, since soft jet signals have an enormous QCD background.

\FIGURE[!t]{\epsfig{file=w1z1_r.eps,width=7.2cm}\quad
\epsfig{file=z2z1_r.eps,width=7.2cm} 
\caption{The $\tw_1 -\tz_1$ (left) 
and $\tz_2 -\tz_1$ (right) mass splittings, 
as functions of $M_1$, for four parameter space slices, 
at $\tan\beta=5,\ 50$ and $\mu>0$ and $\mu <0$, along the lines giving 
$\Omega_{\tz_1} h^2=\Omega_{\rm CDM} h^2$. 
We also include gray points obtained via a random scan 
over the HB/FP region of parameter space.
}
\label{fig:msssplit}}

The $\tw_1 -\tz_1$ ($\tz_2 -\tz_1$) mass gap 
is shown in the left (right) frame of Fig.~\ref{fig:msssplit}
for various values of $\tan\beta$ and signs of $\mu$. 
Positive $\mu$ is beneficial for our purposes, because the mass gap is larger 
than for negative $\mu$, 
and therefore the leptons in the final state are harder. 
A positive value of $\mu$ is currently favored by the discrepancy 
between the measured muon $g-2$ value and the one calculated in the SM using 
$e^+e^-$ data for the diagrams involving hadronic vacuum polarization 
\cite{Passera:2004bj}.

For the rest of this section we restrict ourselves to the following values of 
input parameters: $\tan\beta=50$, $\mu>0$ and $m_t=174.3$ GeV. Unless stated 
otherwise, we will be working along the WMAP favored $\Omega_{\tz_1} h^2=0.11$ 
line in the HB/FP region. 
Along this line $M_1$ and $\mu$ are not independent, and one 
needs to specify only the $M_1$ value.

\subsubsection{Reach via gluino cascade decays}

We use Isajet 7.72 ~\cite{isajet} for the simulation of signal and some of the 
background events at the LHC. A toy detector simulation is employed with 
calorimeter cell size 
$\Delta\eta\times\Delta\phi=0.05\times 0.05$ and $-5<\eta<5$. The hadronic 
energy resolution is taken to be $80\%/\sqrt{E}$ for $|\eta|<2.6$ and 
$100\%/\sqrt{E}$ for $|\eta|>2.6$. The electromagnetic energy resolution 
is assumed to be $3\%/\sqrt{E}$. We use a UA1-like jet finding algorithm 
with jet cone size $R=0.5$ and $p_T^{jet}>25$ GeV. Leptons are considered 
isolated if the visible activity within the cone $\Delta R<0.3$ is 
$\Sigma E_T^{cells}<2$ GeV. The strict isolation criterion helps reduce 
multi-lepton background from heavy quark (especially $t\bar t$) production. 
Leptons ($e$s or $\mu$s) have to satisfy the requirement 
$p_T^{lepton}>10$ GeV. We also require that leptons 
would have $|\eta|<2.5$ and jets would propagate within $|\eta|<3$.
\FIGURE[!t]{\epsfig{file=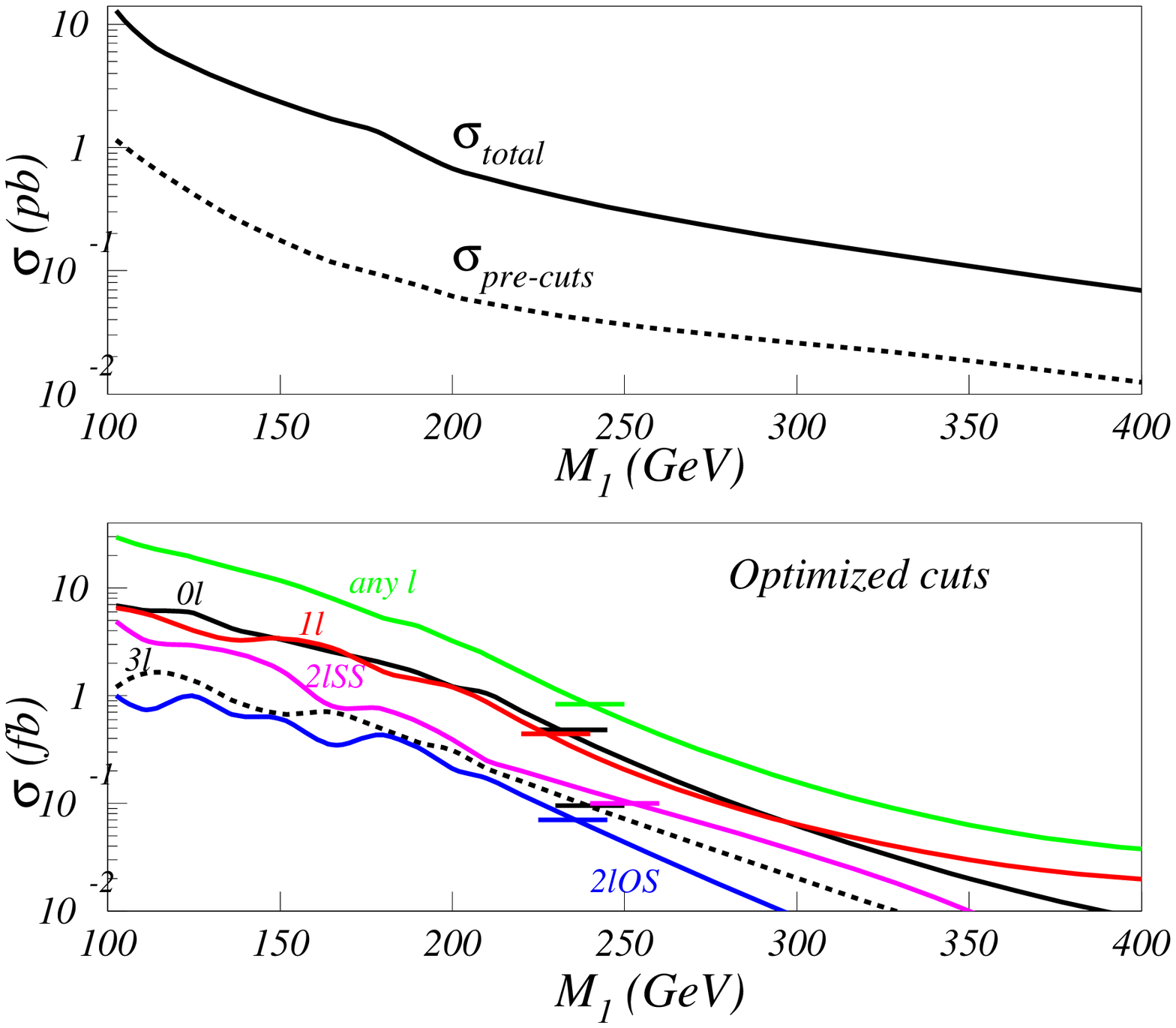,width=10cm}
\caption{
In the upper frame, we show the total cross section and the cross section 
after the pre-cuts ($E_T^{miss}>200$ GeV and at least 2 jets with 
$p_T>40$ GeV and $|\eta |<3$) for SUSY particle production at the
LHC along the $\Omega_{CDM}h^2=0.11$ line, as a function of $M_1$. 
In the lower frame, we show 
LHC cross sections after optimized cuts for various gluino 
cascade decay event topologies and corresponding $5\sigma$ 
discovery limits for 100 
$fb^{-1}$ integrated luminosity along the WMAP favored region. 
LHC can probe to $M_1\sim 250$ GeV, 
which corresponds to $m_{\tilde g}\sim 1.8$ TeV.}
\label{fig:cs_oh_2}}

First, we re-plot the reach of the LHC in Fig.~\ref{fig:cs_oh_2}, using the 
procedure described in \cite{bbbkt}. All events had to 
pass the pre-cuts, which impose the requirement that $E_T^{miss}>200$ GeV 
and there are at 
least 2 jets with $p_T^{jet}>40$ GeV. The definitions of jets and leptons 
can be found in \cite{bbbkt}, as well as the description 
of the cut optimization procedure. We choose the cuts which were found to 
be optimal for the HB/FP region. The events are divided into several classes, 
characterized by the number of leptons in the final state. 
The upper frame of Fig.~\ref{fig:cs_oh_2} 
shows the total supersymmetric particle 
production cross section and the cross section after pre-cuts. 
The lower frame of 
Fig.~\ref{fig:cs_oh_2} shows the signal cross sections after the optimal cuts 
in various channels. The $5\sigma$ discovery reach for 100 $fb^{-1}$ 
of integrated luminosity is shown by short horizontal lines for each channel. 
One can see that the cuts, which are optimized for gluino pair 
production and subsequent cascade decays, provide a reach by the LHC of up to 
$M_1\sim 250$ GeV, corresponding to a value of $m_{\tg}\sim 1.8$ TeV. 
At that point the total sparticle production cross section 
is still sizable: for $M_1=250$ GeV, the total SUSY 
cross section is $\sim 400$ fb. 
This fact motivates us to look next at -ino pair production at the LHC.

\subsubsection{Trilepton production at the LHC in the HB/FP region}
\FIGURE[!t]{\mbox{\hspace{-1cm}
\epsfig{file=m1__neutneut.eps,width=8cm}\quad
\epsfig{file=m1__neutchar1.eps,width=8cm}}
\caption{The leading order production cross sections at the LHC 
for all possible pairs of neutralinos (left), 
and of a chargino plus one of the two lightest neutralinos (right), 
along the line at $\Omega_{\tz_1} h^2=0.11$.}
\label{fig:xsec1}}
\FIGURE[!t]{\mbox{\hspace{-1cm}
\epsfig{file=m1__neutchar2.eps,width=8cm}\quad
\epsfig{file=m1__charchar.eps,width=8cm}}
\caption{The leading order production cross sections at the LHC 
for all possible pairs of a chargino plus one of the two heaviest 
neutralinos (left), and of two charginos (right), 
along the line at $\Omega_{\tz_1} h^2=0.11$.}
\label{fig:xsec2}}

Total production cross sections for neutralino pair production at the LHC 
along the line of constant $\Omega_{\tz_1} h^2 =0.11$ are 
shown versus $M_1$ in the left frame of Fig.~\ref{fig:xsec1}. 
The right frame of 
Fig.~\ref{fig:xsec1} shows total cross sections for associated 
production of the lightest or next-to-lightest neutralino with a chargino. 
Similarly, we present the associated production of 3rd or 4th lightest 
neutralino with a chargino in the left frame of Fig.~\ref{fig:xsec2} and 
the chargino pair production cross section in the right frame of 
Fig.~\ref{fig:xsec2}.
Of all the -ino production cross sections, $\tw_1\tz_1$, $\tw_1\tz_2$ and
$\tw_1^+\tw_1^-$ production are generally the largest.

The gluino pair production cross section is presented in 
Fig.~\ref{fig:gluinocs} versus $M_1$. 
We have also shown the total -ino production cross section for comparison. 
Assuming 100 $fb^{-1}$ integrated luminosity, LHC would produce less than 
10 gluino pairs for $M_1\sim 300$ GeV, while $~\sim 10^4$ -ino 
pairs would be produced.

\FIGURE[!t]{\epsfig{file=gluinogluino.eps,width=8cm}
\caption{The gluino pair production cross section, 
along the good-relic-abundance slice at $\tan\beta=50$ and $mu>0$, 
compared to the total neutralino-chargino production cross section. 
On the upper x-axis we also indicate the reference gluino mass.}
\label{fig:gluinocs}}

Given the relative total production cross section rates, it might be beneficial
to examine -ino pair production signals in the HB/FP region, as well
as gluino pair signals.
Single lepton signals from $\tw_1\tz_1$ where $\tw_1\to \ell\nu_\ell \tz_1$ 
will be buried under an immense background from direct $W$ boson production.
Likewise, dilepton production from reactions such as $\tw_1^+\tw_1^-$
production will be buried beneath large backgrounds from $W^+W^-$ and
$t\bar{t}$ production.
Four lepton signals from reactions such as $\tz_2\tz_2$ 
production would be very distinctive and it is possible to find the 
cuts which reduce the SM background. However, our preliminary 
analysis found that the 4 lepton signal rates fall very quickly with 
increasing $M_1$, and this channel did not provide any additional reach 
compared to the optimized cuts for gluino pair production. Thus we were led 
to consider the clean trilepton signature at the LHC in more detail\cite{lhc3l}. 
This signal provides 
the best reach in mSUGRA at the Tevatron ~\cite{bkt,trilep,bk}, due to 
$\tilde W_1\tilde Z_2$ production and subsequent decays 
$\tilde Z_2\rightarrow l\bar l\tilde Z_1$ and 
$\tilde W_1\rightarrow l\bar\nu_l\tilde Z_1$.

Let us first examine the dominant production processes, which could produce 
3 or more leptons in the final state, at $M_1=250$ GeV, where the reach due 
to optimized cuts peters out:
\begin{eqnarray}
\sigma(\tilde W_1\tilde Z_2)&=&70.8 fb,\label{w1z2}\\
\sigma(\tilde W_1\tilde Z_3)&=&76.8 fb,\label{w1z3}\\
\sigma(\tilde Z_2\tilde Z_3)&=&28.0 fb,\label{z2z2}\\
\sigma(\tilde W_2\tilde W_2)&=&17.2 fb,\label{w2w2}\\
\sigma(\tilde W_2\tilde Z_4)&=&36.4 fb.\label{w2z4}
\end{eqnarray}
The relevant branching fractions are:
\begin{eqnarray*}
BF(\tilde W_1\rightarrow \tilde Z_1 l\bar\nu_l)&=&0.22\\
BF(\tilde Z_2\rightarrow \tilde Z_1 l\bar l)&=&0.07\\
BF(\tilde Z_3\rightarrow \tilde Z_1 l\bar l)&=&0.07\\
BF(\tilde W_2\rightarrow \tilde Z_2 W)&=&0.27,\ \
BF(\tilde W_2\rightarrow \tilde Z_3 W)=0.25, \\
BF(\tilde W_2\rightarrow \tilde W_1 Z)&=&0.27,\ \
BF(\tilde W_2\rightarrow \tilde W_1 h)=0.2\\
BF(\tilde Z_4\rightarrow \tilde W_1 W)&=&0.58,\ \
BF(\tilde Z_4\rightarrow \tilde Z_2 Z)=0.02, \\
BF(\tilde Z_4\rightarrow \tilde Z_3 Z)&=&0.21,\ \
BF(\tilde Z_4\rightarrow \tilde Z_2 h)=0.17,
\end{eqnarray*}
where $l$ stands for either $e$ or $\mu$.
It is now possible to estimate the maximal possible 3 lepton event rate 
\it before any cuts \rm for 100 $fb^{-1}$ integrated luminosity at the LHC. We 
do that only for the processes 
(\ref{w1z2}), (\ref{w1z3}) and (\ref{z2z2}) here:
\begin{eqnarray*}
N_{3l}(\tilde W_1\tilde Z_2)&=&109,\\
N_{3l}(\tilde W_1\tilde Z_3)&=&118,\\
N_{3l}(\tilde Z_2\tilde Z_3)&=&14.
\end{eqnarray*}

Next we proceed to the fast 
simulation. The dominant backgrounds for the 
clean trilepton signature are $t\bar t$, 
$WZ$ and $ZZ$ production.
When evaluating $WZ$ production, it has been shown to be of crucial
importance to evaluate the full $q\bar{q}'\to \ell\nu_\ell \ell'\bar{\ell}'$
background, which includes off-shell $W^\ast Z^\ast$ and 
$W^\ast\gamma^\ast$ production, as well as other diagrams\cite{trilep}.
We have used Isajet 7.72~\cite{isajet} 
to calculate the $t\bar t$, $ZZ$ and $WZ$ with $Z\to\tau\bar{\tau}$ 
backgrounds. For off-shell 
$W^\ast Z^\ast$ and $W^\ast\gamma^\ast$ background calculation, we have 
employed an exact tree level evaluation of the $2\to 4$ processes 
using Madgraph\cite{madgraph} at the parton level.
A similar calculation has been performed in Refs.~\cite{trilep}, 
where soft lepton $p_T$ cuts were invoked\cite{bk}to try to maximize the
signal when leptons originate from $\tau $ decays. In our case, in the LHC
environment, we will require 
three isolated leptons each with $p_T^{lepton}>10$ GeV and
$|\eta|<2.5$ throughout our analysis.
In Ref. \cite{trilep}, also
{\it i}. an invariant mass cut of $m(\ell'\bar{\ell}')<81$ GeV was invoked to reduce
BG from the on-shell $Z$ boson contribution, {\it ii}. $m(\ell'\bar{\ell}')>20$ GeV was used
to reduce BG from the photon pole, and {\it iii}. a transverse mass veto of
$65\ {\rm GeV}<M_T(\ell ,\nu_{\ell})<80$ GeV was used to reduce BG from on-shell
$W$ contributions. Finally, $\eslt >25$ GeV was required. These cuts, 
dubbed SC2 by the Tevatron Run2 study group, will be invoked here, 
along with the somewhat stronger lepton $p_T$ cuts.
The BG rates after the SC2 cuts are listed in Table \ref{tab:3l}.
The $5\sigma$ signal for cuts SC2 assuming 100 fb$^{-1}$ of integrated
luminosity is thus $1.38$ fb.
In Table \ref{tab:3l}, we also list a signal point with $M_1=110$ GeV, 
where gluino pair production is still large and the signal is very robust. 
\begin{table}[!b]
\begin{center}
\begin{tabular}{l@{\hspace{10mm}}cc}
\hline
\hline
\\[-2mm]
Process  &  $\sigma (tot)$ (fb) &  $\sigma (SC2)$ (fb) \\[1mm]
\hline
\vspace{.1cm}
$t\bar t$ & $5.3\times 10^5$ & 4.5 \\
$WZ\ (W\to\ell\nu ,\ Z\to \tau\bar{\tau})$ & 238 & 0.79 \\
$ZZ\ (Z\to \ell\bar{\ell},\ \nu\bar{\nu})$ & 758.3 & 0.36 \\
 $W^\ast Z^\ast, W^\ast\gamma^\ast$ & --- & 2.0 \\
\hline
Total & --- & 7.65 \\
\hline
\hline
Case study at $M_1=110$ GeV & 7796 & 13.1 \\
\hline
\end{tabular}
\end{center}
\label{tab:3l}
\end{table}
%

\FIGURE[!t]{\epsfig{file=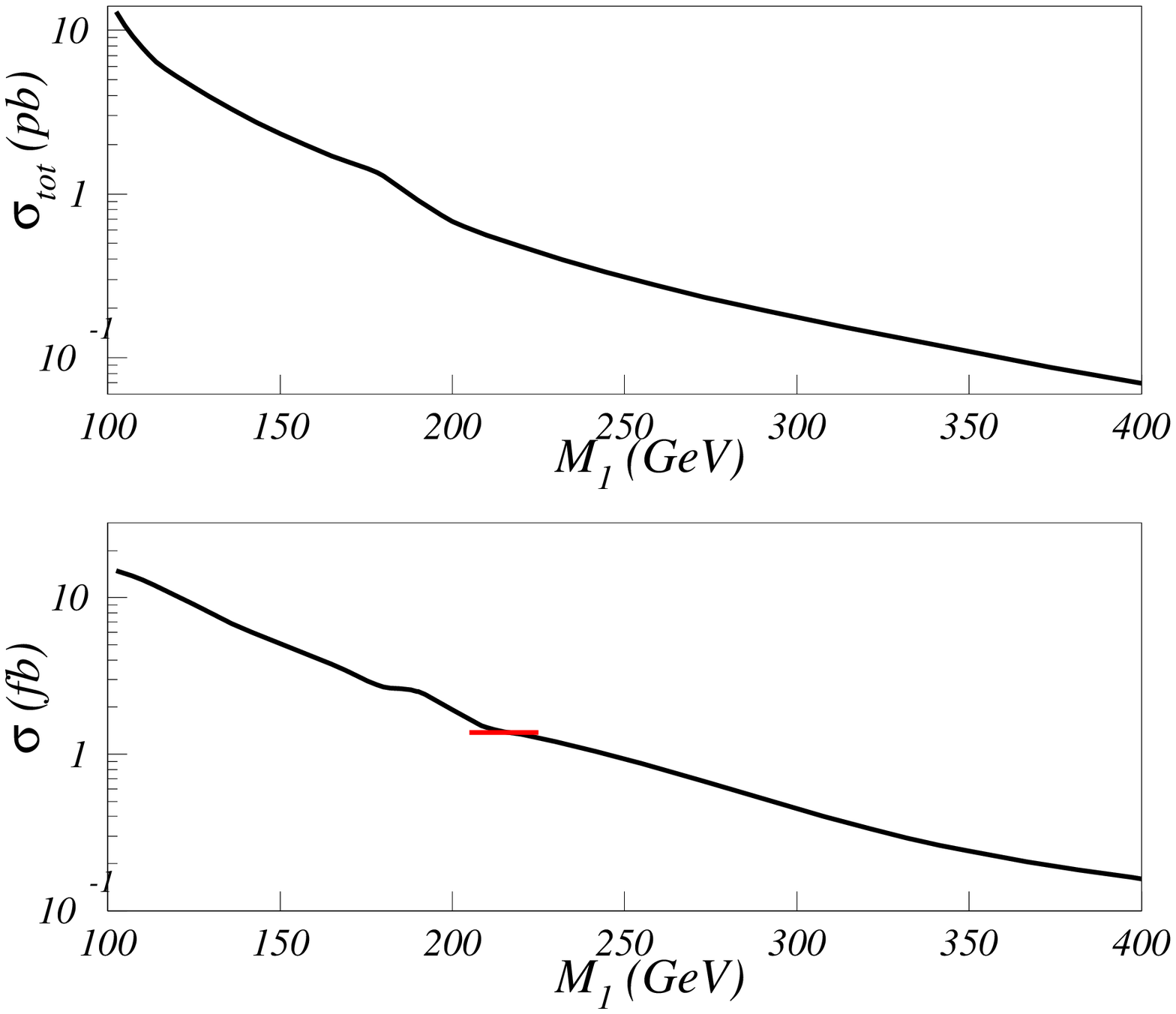,width=9cm}
\caption{
(Upper) Total cross section for $pp\to$ SUSY particles at the CERN LHC 
along the $\Omega_{\tz_1} h^2=0.11$ line, as a function of $M_1$.
(Lower): LHC clean trilepton cross section after cuts SC2 
and corresponding $5\sigma$ discovery limits for 100 
$fb^{-1}$ integrated luminosity along the 
$\Omega_{\tz_1} h^2=0.11$ line as a function of $M_1$. 
}
\label{fig:cs_oh_sc2}}

In Fig. \ref{fig:cs_oh_sc2}, upper frame, we show the 
total $pp\to\tw_1\tz_2$ cross section, and in the lower frame, the clean trilepton cross section
after cuts SC2 along the line of $\Omega_{\tz_1}h^2 =0.11$ as a 
function of $M_1$. Also shown by the horizontal mark is the
$5\sigma$ limit for 100 fb$^{-1}$ of integrated luminosity. We see
that the CERN LHC reach for clean trileptons is possible out to 
$M_1\sim 220$ GeV, which is, in fact, comparable to the reach via
conventional cascade decay signatures shown in Fig. \ref{fig:cs_oh_2}, left.

While the results of Fig. \ref{fig:cs_oh_sc2}, left, are valid along the line
$\Omega_{\tz_1} h^2 =0.11$, it is also possible that $\Omega_{\tz_1} h^2 <0.11$
in scenarios with mixed dark matter. In this case, values of $\mu$
smaller than those used in Fig. \ref{fig:cs_oh_sc2}, left, are possible.
As $\mu$ becomes smaller, then the $\tz_1$ becomes even more
higgsino-like, and the relic density drops. The masses
$m_{\tz_{1,2}}$ and $m_{\tw_1}$ drop as well, as does the mass gap
$m_{\tz_2}-m_{\tz_1}$. 
The situation is illustrated in Fig.~\ref{fig:cs_oh_sc2}, right, for the case of 
$M_1=220$ GeV, where various -ino masses are plotted versus a variable $\mu$
parameter.
\FIGURE[!t]{\mbox{\hspace{-1cm}
\epsfig{file=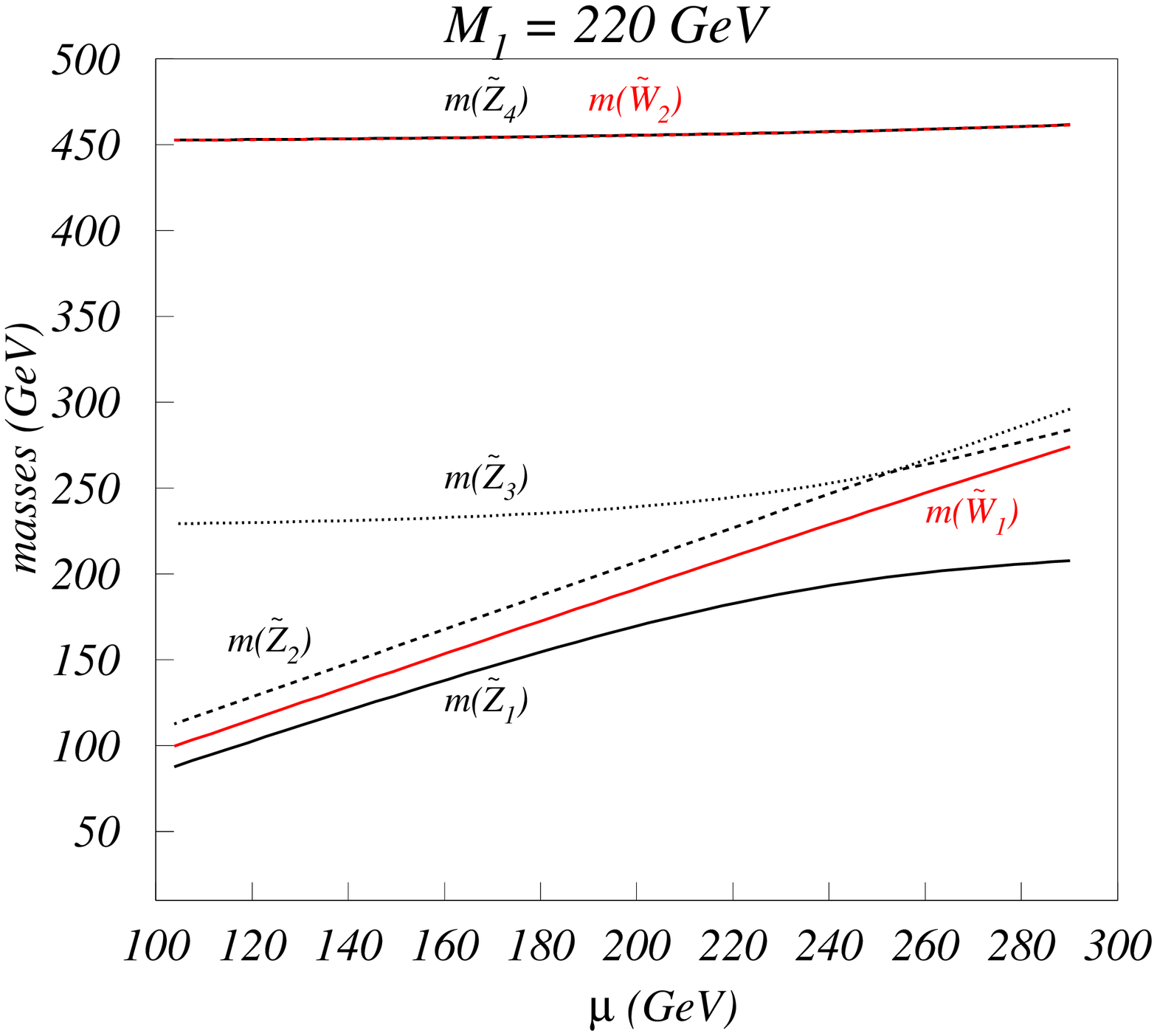,width=8.5cm}
\epsfig{file=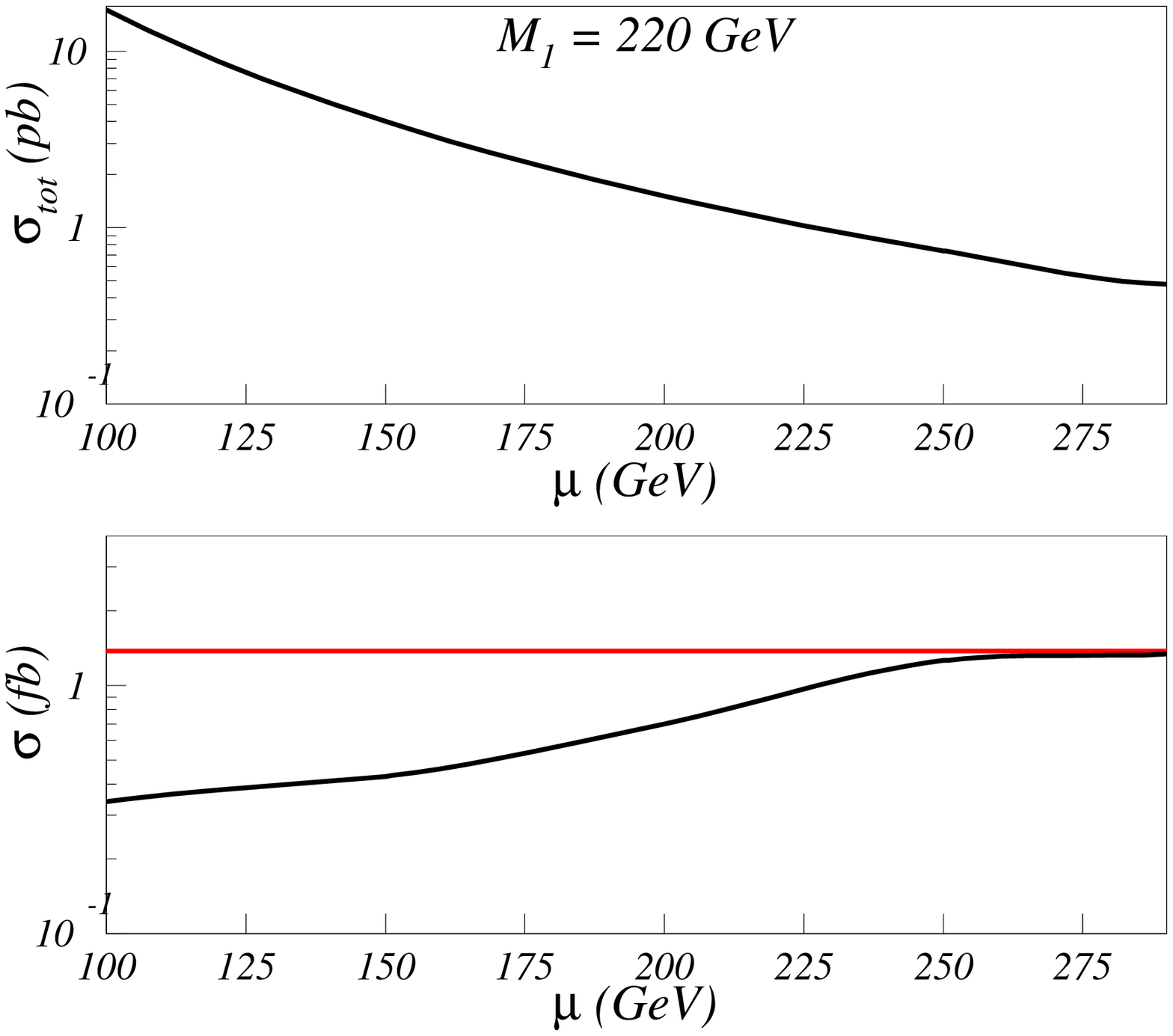,width=8.5cm} }
\caption{(Left): Masses of lighter gauginos for a fixed $M_1=220$ GeV 
along the 
region with $\Omega_{\tz_1} h^2\le 0.11$.
(Right): LHC total SUSY particle cross section (upper) and 
clean trilepton cross section after cuts SC2 (lower) 
along with corresponding $5\sigma$ 
discovery limits for 100 
$fb^{-1}$ integrated luminosity for a fixed $M_1=220$ GeV along the 
line of varying $\mu$.}
\label{fig:cs_m1fix}}
\FIGURE[!t]{\epsfig{file=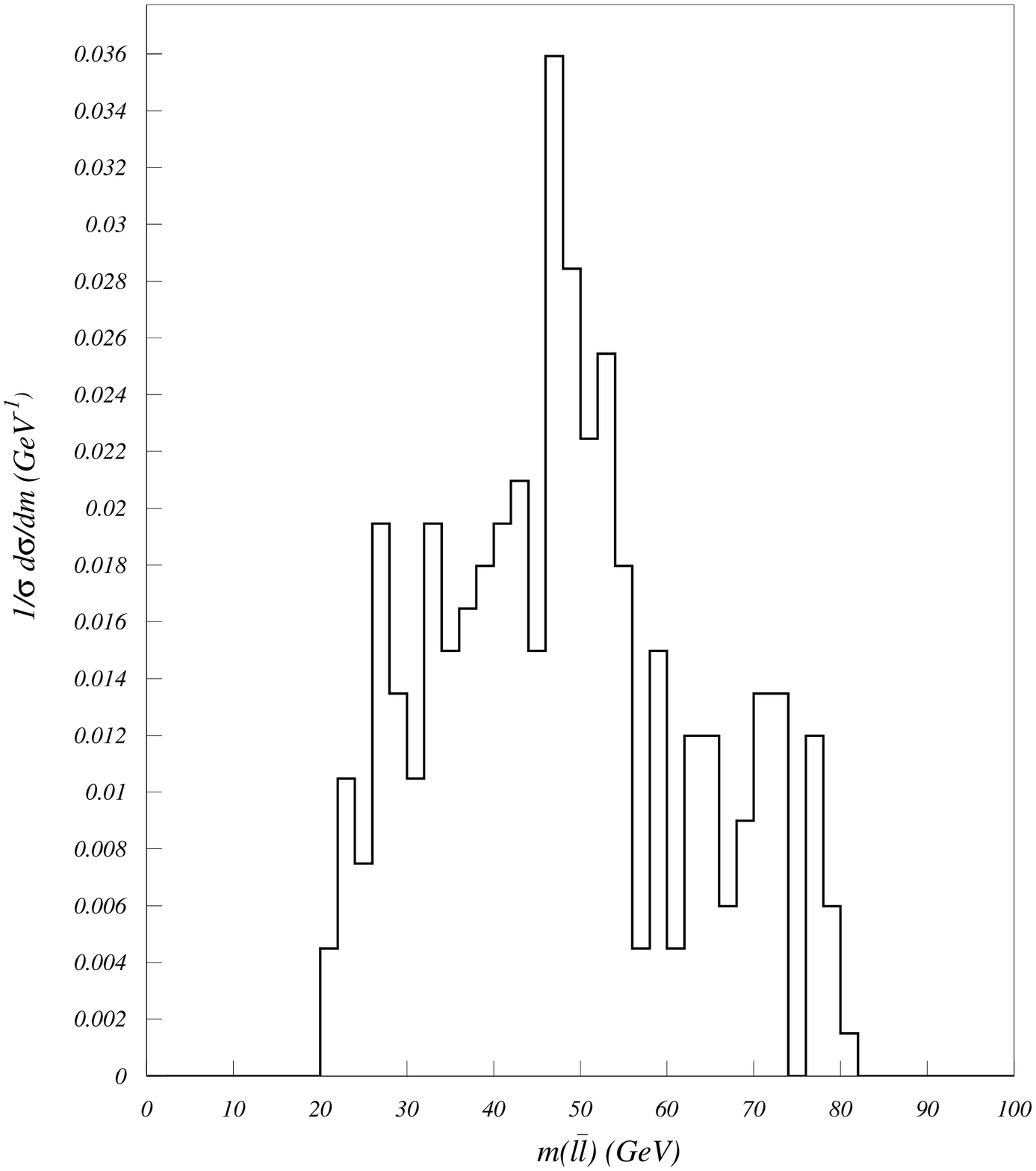,width=7.7cm}
\caption{Distribution in opposite sign/same flavor dilepton invariant mass
from clean trilepton events at the CERN LHC for $M_1=110$ GeV along the
line of $\Omega_{\tz_1} h^2 =0.11$.}
\label{fig:mll}}

Thus, as $\mu$ drops to smaller values, many
of the -ino production cross sections rise. However, the trilepton
energy and momentum distributions will diminish, in part due to the 
reduced parent particle masses, and in part due to the reduced 
sparticle decay mass gaps, which lead to smaller energy release
in the chargino and neutralino decays. Thus, as the value of $\mu$ 
is reduced, production cross sections 
increase, while detection efficiency decreases. 
The trilepton cross section after cuts SC2 is shown versus $\mu$ for
fixed $M_1=220$ GeV in Fig. \ref{fig:cs_m1fix}, right.
Here we see that at large $\mu$ values, the trilepton cross section 
after cuts is at the edge of observability. However, as $\mu$ decreases,
the reduced detection efficiency wins out over the increasing 
production cross section (shown in the lower frame of Fig. \ref{fig:cs_m1fix}), left,
resulting in an overall diminished trilepton cross section. Thus, the trilepton
cross section is actually maximal along the line $\Omega_{\tz_1} h^2 =0.11$,
and diminishes in regions where $\Omega_{\tz_1} h^2 <0.11$.

In Fig. \ref{fig:mll}, we show the opposite-sign/same flavor dilepton 
invariant mass distribution from clean trilepton events at the CERN LHC for
the case where $M_1=110$ GeV. It is important to note {\it two}
distinct mass edges in the plot. The first occurs from the kinematical edge
from $\tz_2\to\tz_1 f\bar{f}$ decay, and occurs at 
$m_{\tz_2}-m_{\tz_1}=56.4$ GeV. The second comes from $\tz_3\to\tz_1 f\bar{f}$
decay, and occurs at $m_{\tz_3}-m_{\tz_1}=82.1$ GeV. The latter mass gap 
is close enough to the $Z$-pole that decay matrix element effect skews the
invariant mass towards the high end of the range.

\subsection{Prospects for sparticle detection at the ILC}

The proposed International Linear Collider is projected to operate
initially at $\sqrt{s}\sim 0.5$ TeV with an integrated luminosity of
$\sim 100$ fb$^{-1}$. In its later stages, the CM energy should be upgraded
to $\sqrt{s}\simeq 1$ TeV.
In Ref. \cite{bbktnew}, it has been shown that the reach of a linear collider
can exceed that of the CERN LHC in the HB/FP region. This is because $|\mu |$
is always small in the HB/FP region, which forces detectable charginos and neutralinos--
which should be readily accessible to $e^+e^-$ colliders if the beam 
energy is sufficiently high-- to be relatively light, 
even if the scalars and the gluino are relatively heavy.

In Fig. \ref{fig:ILC}, we show total production cross sections for
various -ino pair production reactions versus $M_1$ along the line
of constant $\Omega_{\tz_1} h^2=0.11$. The left-hand frame is for a 
$\sqrt{s}=0.5$ TeV collider, while the right-hand frame is for
a $1$ TeV collider.
By examining frame {\it a}), we see that for $M_1\sim 100-200$ GeV,
the -ino production reactions are dominated by chargino pair production,
although a variety of other reactions such as $\tz_2\tz_3$, $\tz_1\tz_3$
and $\tw_1\tw_2$ may also be accessible. The mass spectrum shown earlier 
in Fig. \ref{fig:splitting} shows that $m_{\tw_2}$ and $m_{\tz_4}$ are of order
$M_2$, and so should be heavy enough that two-body decays are accessible.
However, $\tw_1$, $\tz_2$ and $\tz_3$ have large higgsino components 
and are relatively light; 
they should typically decay via three-body modes, where the 
branching fractions are dominated by the $W$ or $Z$ boson propagators
(since scalars are assumed quite heavy). Thus, the decays of 
$\tz_2$ and $\tz_3$ should be very similar in that $\tz_2\to\tz_1 f\bar{f}$
and $\tz_3\to\tz_1 f\bar{f}$, aside from the size of the $\tz_3 -\tz_1$
vs. the $\tz_2 -\tz_1$ mass gaps. We see that the ultimate reach of the
$\sqrt{s}=0.5$ TeV machine is determined by the $\tz_1\tz_3$ cross section.
$\tz_1\tz_2$ production is more favorable kinematically, but has a 
lower total cross section due to a suppressed $Z\tz_1\tz_2$ coupling.
The reach of a 0.5 TeV ILC along the $\Omega_{\tz_1} h^2 =0.11$ line is out to
$M_1\sim 220$ GeV (corresponding to $m_{\tg}\sim 1650$ GeV), 
and is slightly below the reach of the CERN LHC.
\FIGURE[!t]{\mbox{\hspace{-0.5cm}\epsfig{file=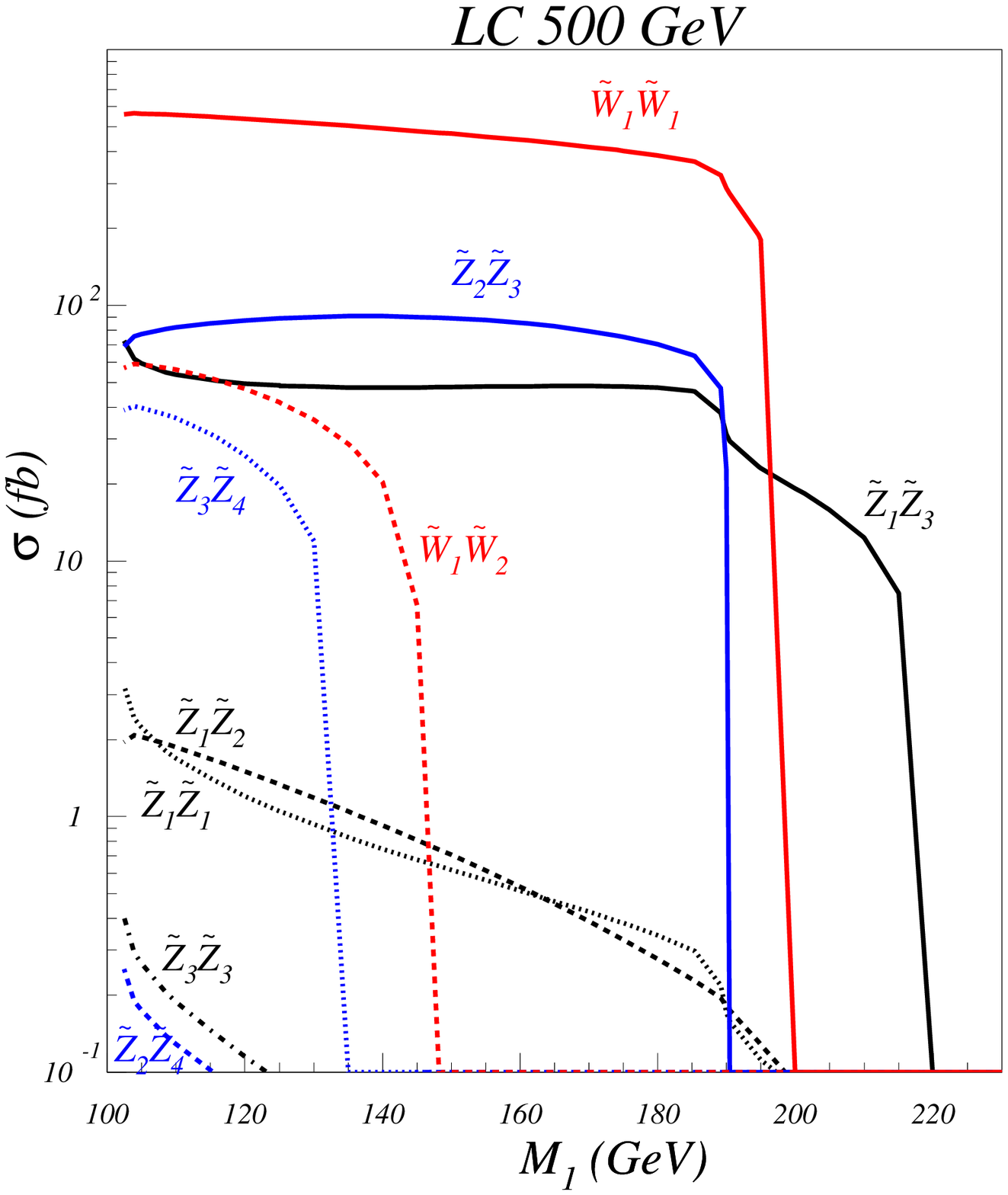,width=7.5cm}\quad
\epsfig{file=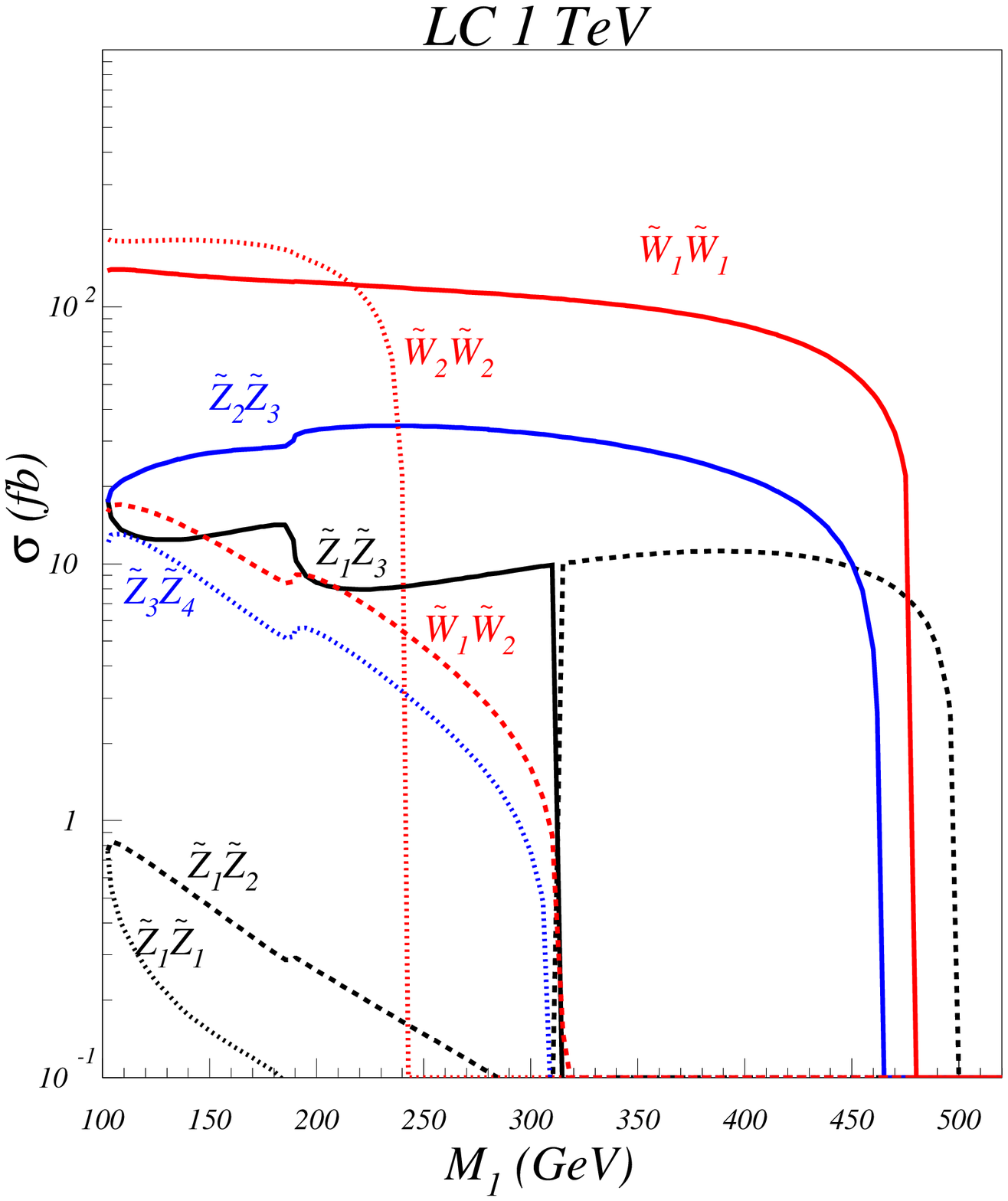,width=7.5cm}}
\caption{Chargino and neutralino production cross sections 
at the ILC with $\sqrt{s}=0.5$  TeV
(left) and 1 TeV (right).}
\label{fig:ILC}}

In frame {\it b}), for a $\sqrt{s}=1$ TeV machine, we again see that
$\tw_1^+\tw_1^-$ production is dominant over most of the range of $M_1$,
while again a variety of other -ino pair production reactions should in general 
be present. In this case, the ultimate reach is determined by the $\tz_1\tz_2$
production reaction, and extends out to $M_1\sim 500$ GeV (corresponding to
$m_{\tg}\sim 3.5$ TeV), far beyond the reach of the CERN LHC.

\subsection{Gluino lifetime in the HB/FP region}

The gluino decay width for $\widetilde g\rightarrow q\overline q \widetilde{\tz_1}$ can be expressed as
\begin{equation}
\Gamma_{\tg}\sim c^2\ \frac{\alpha_s\alpha}{48\pi}\  \frac{m^5_{\widetilde g}}{m^4_{\widetilde q}},
\end{equation}
where $c$ represents a suitable combination of the neutralino-squark-quark couplings. The gluino lifetime, taking into account that the factor $\frac{\alpha_s\alpha}{48\pi}\sim6.6\times10^{-6}$, evaluating $\alpha_s$ and $\alpha$ at $M_Z$, can thus be cast as
\begin{equation}
\tau\sim\frac{10^{-19}}{c^2}\frac{m^4_{\widetilde q}}{m^5_{\widetilde g}}\ {\rm sec}
\end{equation}
Conservatively assuming that $c<0.1$, we obtain
\begin{equation}
\tau<10^{-17}\frac{m^4_{\widetilde q}}{m^5_{\widetilde g}}\ {\rm sec}
\end{equation}
In the case of minimal supergravity, we can draw the following general upper limit on the gluino lifetime. In mSUGRA, we always have $m_{\widetilde g}\gtrsim2.5\times m_{1/2}$, and, in the focus point region, where the gluino lifetime is maximal, 
we can safely take $m_{\widetilde q}\sim m_0\lesssim100\times M_{1/2}$. We therefore have
\begin{equation}
\tau<10^{-11}\left(\frac{\rm GeV}{m_{1/2}}\right)\ {\rm sec}.
\end{equation}
Since in the focus point region $m_{1/2}\gtrsim100$ GeV, we find that, in minimal supergravity, $\tau>10^{-13}$ sec. 
In order to detect a displaced vertex, the lifetime of a quasi-stable particle should be at least larger than 
$\tau_{\rm disp}\sim 10^{-12}$ sec. 
This therefore entails that {\em in mSUGRA gluinos are never ``stable'' inside a detector}, 
which means that {\em if a ``stable'' gluino , or a displaced gluino vertex is detected, the underlying SUSY theory cannot be mSUGRA}. 

\FIGURE[!t]{\epsfig{file=gluino_lifetime.eps,width=9.5cm} 
\caption{The gluino lifetime as a function of the GUT-scale 
universal scalar mass $m_0$, along mSUGRA slices at 
$\tan\beta=50$, $\mu>0$, $A_0=0$ for two fixed values of 
$m_{1/2}=500$ and 1000 GeV.}
\label{fig:gluino}}

The largest gluino lifetimes are obtained in the focus point region at low values of $m_{1/2}$ and of $\tan\beta$, using a large top mass input. 
The spread in the gluino lifetime within mSUGRA can be significant, ranging from a minimum in the stau coannihilation region, where the squarks are the lightest possible, to a maximum in the HB/FP region. 

In the stau coannihilation region, we numerically find that $c^2\sim0.5$, which gives an estimated gluino lifetime between $10^{-24}$ and $10^{-26}$ sec, depending on the value of $\tan\beta$. The same range of gluino lifetimes is expected in the bulk and funnel regions. In the 
HB/FP region, one instead always obtains gluino lifetimes larger than $10^{-23} $ sec. Since the gluino hadronizes if $\tau>\tau_{\rm had}\sim 6.6\times 10^{-25}{\ \rm sec}/(m_\pi/{\rm GeV})\sim10^{-23}$ sec, this means that the cosmologically viable parameter space of mSUGRA is split, from the point of view of gluino lifetimes, in a {\em non-hadronizing gluino branch} (coannihilation, funnel and bulk regions) and in a {\em hadronizing gluino branch} (HB/FP region). 
A hadronizing gluino gives rise to {\it i}). additional hadrons through fragmentation, {\it ii}). smeared out spin correlations and
{\it iii}). reduced energy in daughter particle energy distributions. All of these effects will, however, be very difficult
to determine experimentally in the environment of an LHC detector. 


\section{Conclusions}
\label{sec:conclusions}

The HB/FP region of SUGRA models is very compelling from both a theoretical
and phenomenological viewpoint. The multi-TeV scalars act to suppress
unwanted FCNCs, CP-violating dipole moments and proton decay, 
while maintaining low fine-tuning. In addition, the HB/FP region is 
entirely consistent with the WMAP CDM constraint, due to the presence of
a mixed bino-higgsino LSP. Previously, many investigations of phenomenology 
in the HB/FP region took place in the mSUGRA ($m_0,\ m_{1/2}$) plane.
Calculations of HB/FP phenomenology 
in terms of these GUT scale parameters are plagued by numerical and parametric
uncertainties. We propose here, instead, a presentation of HB/FP phenomenology 
in terms of weak scale ($M_1,\ \mu $) parameters, wherein {\it i}). 
the {\it entire} HB/FP region can be displayed, including the $\mu \to 0$ region, 
and {\it ii}). the parameter plane is much more stable against variations 
in other parameters such as $m_t$ and $\tan\beta$.

We present the HB/FP in the ($M_1,\ \mu $) plane, and show regions that 
are allowed and disallowed by theory, collider and astrophysical constraints.
We also show prospects for future DM  searches in the ($M_1,\ \mu$) plane.
Prospects for exploration of the WMAP allowed portion of the HB/FP plane
are excellent for Stage 3 direct dark matter detectors, for
indirect detection via neutrino telescopes, and for antideuteron searches
via the GAPS experiment. Prospects may also be good for $\gamma$ ray searches 
from neutralino annihilation in the galactic core, but these estimates depend 
sensitively on the assumed dark matter halo distribution near the 
galactic center.

Regarding collider searches in the WMAP allowed portion of the HB/FP region,
we find that the LHC, with 100 fb$^{-1}$ of data, should be able to probe
to $m_{\tg}\sim 1.8$ TeV via conventional gluino cascade decay signatures.
We also studied the reach of the LHC for SUSY in the clean trilepton 
channel, including backgrounds from $W^*Z^*$ and $W^*\gamma^*$
production. 
In this case, the LHC reach for 100 fb$^{-1}$ is slightly smaller, 
out to $m_{\tg}\sim 1.65$ TeV. A linear
$e^+e^-$ collider will have a reach in the HB/FP region determined by
$\tz_1\tz_2$ and $\tz_1\tz_3$ production. A $\sqrt{s}=0.5$ TeV LC
will have a reach to $m_{\tg}\sim 1.65$ TeV, while a $\sqrt{s}=1$ TeV
LC will have a reach beyond that of the CERN LHC, 
out to $m_{\tg}\sim 3.5$ TeV.

\acknowledgments
 
We thank Frank Paige for useful discussions.
H.B. and S.P. were supported in part by the U.S. Department of Energy
under contract number DE-FG02-97ER41022. P.U. was supported in part by the RTN 
project under grant HPRN-CT-2000-00152 and by the Italian INFN under the
project ``Fisica Astroparticellare''. T. K. was supported in part by the 
U.S. Department of Energy under contract No. DE-AC02-98CH10886.

	
%

\end{document}